\documentclass{article}
\pdfoutput=1
\usepackage[T1]{fontenc}
\usepackage{lmodern}
\usepackage{slantsc}
\usepackage[protrusion=true,expansion=true]{microtype}
\usepackage{breakcites}
\usepackage{amsmath,amssymb,amsfonts,amsthm}
\usepackage{subcaption}
\usepackage{graphicx}
\usepackage{fullpage}
\usepackage{setspace}
\usepackage[numbers]{natbib}
\usepackage{booktabs}
\usepackage[backref=page]{hyperref}
\usepackage{color}
\usepackage{wrapfig}
\usepackage{tikz}
\usetikzlibrary{decorations.pathreplacing}
\usepackage{algorithm}
\usepackage[noend]{algpseudocode}
\usepackage[framemethod=tikz]{mdframed}
\usepackage{xspace}
\usepackage{pgfplots}
\usepackage{framed}
\usepackage{thmtools}
\usepackage{thm-restate}
\usepackage{tabu}
\usepackage{fancyhdr}
\usepackage{enumitem}
\pgfplotsset{compat=1.5}

\newtheorem{theorem}{Theorem}[section]

\newtheorem{lemma}[theorem]{Lemma}

\newtheorem{definition}[theorem]{Definition}

\newenvironment{proofof}[1]{\begin{trivlist} \item {\bf Proof
#1:~~}}
  {\qed\end{trivlist}}

\newcommand{\namedref}[2]{\hyperref[#2]{#1~\ref*{#2}}}
\newcommand{\thmlab}[1]{\label{thm:#1}}
\newcommand{\thmref}[1]{\namedref{Theorem}{thm:#1}}
\newcommand{\lemlab}[1]{\label{lem:#1}}
\newcommand{\lemref}[1]{\namedref{Lemma}{lem:#1}}

\newcommand{\seclab}[1]{\label{sec:#1}}
\newcommand{\secref}[1]{\namedref{Section}{sec:#1}}

\newcommand{\figlab}[1]{\label{fig:#1}}
\newcommand{\figref}[1]{\namedref{Figure}{fig:#1}}
\newcommand{\alglab}[1]{\label{alg:#1}}
\renewcommand{\algref}[1]{\namedref{Algorithm}{alg:#1}}
\newcommand{\tablelab}[1]{\label{tab:#1}}
\newcommand{\tableref}[1]{\namedref{Table}{tab:#1}}
\newcommand{\deflab}[1]{\label{def:#1}}
\newcommand{\defref}[1]{\namedref{Definition}{def:#1}}

\newcommand{\PPr}[1]{\ensuremath{\mathbf{Pr}\left[#1\right]}}

\newcommand{\Ex}[1]{\ensuremath{\mathbb{E}\left[#1\right]}}

\renewcommand{\O}[1]{\ensuremath{\mathcal{O}\left(#1\right)}}
\newcommand{\tO}[1]{\ensuremath{\tilde{\mathcal{O}}\left(#1\right)}}
\newcommand{\eps}{\varepsilon}

\def \calA    {\mdef{\mathcal{A}}}
\def \calB    {\mdef{\mathcal{B}}}
\def \calC    {\mdef{\mathcal{C}}}

\def \calE    {\mdef{\mathcal{E}}}
\def \calF    {\mdef{\mathcal{F}}}

\def \calL    {\mdef{\mathcal{L}}}

\def \calR    {\mdef{\mathcal{R}}}
\def \calS    {\mdef{\mathcal{S}}}
\def \calT    {\mdef{\mathcal{T}}}

\def \calX    {\mdef{\mathcal{X}}}

\newcommand{\mdef}[1]{{\ensuremath{#1}}\xspace}  

\DeclareMathOperator*{\dist}{dist}

\DeclareMathOperator*{\poly}{poly}

\newcommand{\set}[1]{\mdef{\left\{#1\right\}}}                        

\newcommand{\ignore}[1]{}

\newif\ifnotes\notestrue 
\ifnotes
\newcommand{\samson}[1]{\textcolor{blue}{{\bf (Samson:} {#1}{\bf ) }} 
\marginpar{\tiny\bf
             \begin{minipage}[t]{0.5in}
               \raggedright S:
            \end{minipage}}}
\newcommand{\david}[1]{\textcolor{purple}{{\bf (David:} {#1}{\bf ) }} \marginpar{\tiny\bf
             \begin{minipage}[t]{0.5in}
               \raggedright D:
            \end{minipage}}} 
\newcommand{\shenghao}[1]{\textcolor{red}{{\bf (Shenghao:} {#1}{\bf ) }} 
\marginpar{\tiny\bf
             \begin{minipage}[t]{0.5in}
               \raggedright S:
            \end{minipage}}}
\else
\newcommand{\samson}[1]{}
\newcommand{\david}[1]{}
\newcommand{\shenghao}[1]{}
\fi

\makeatletter
\renewcommand*{\@fnsymbol}[1]{\textcolor{mahogany}{\ensuremath{\ifcase#1\or *\or \dagger\or \ddagger\or
 \mathsection\or \triangledown\or \mathparagraph\or \|\or **\or \dagger\dagger
   \or \ddagger\ddagger \else\@ctrerr\fi}}}
\makeatother

\providecommand{\email}[1]{\href{mailto:#1}{\nolinkurl{#1}\xspace}}

\definecolor{mahogany}{rgb}{0.75, 0.25, 0.0}
\definecolor{darkblue}{rgb}{0.0, 0.0, 0.55}
\definecolor{darkpastelgreen}{rgb}{0.01, 0.75, 0.24}
\definecolor{darkgreen}{rgb}{0.0, 0.2, 0.13}
\definecolor{darkgoldenrod}{rgb}{0.72, 0.53, 0.04}
\definecolor{darkred}{rgb}{0.55, 0.0, 0.0}
\definecolor{forestgreenweb}{rgb}{0.13, 0.55, 0.13}
\definecolor{greencss}{rgb}{0.0, 0.5, 0.0}
\definecolor{bleudefrance}{rgb}{0.19, 0.55, 0.91}

\hypersetup{
     colorlinks   = true,
     citecolor    = mahogany,
	 linkcolor	  = greencss,
     urlcolor = mahogany
}

\fancypagestyle{pg}
{
\lhead{}
\rhead{}
\cfoot{--\ \thepage\ --}

}

\title{Fast Diverse Nearest Neighbor Search}

\author{
Justin Y. Chen\\
MIT\\
\email{justc@mit.edu}
\and
Soham Nagawanshi\\
Texas A\&M University\\
\email{ndsoham@tamu.edu}
\and
Shenghao Xie\\
Carnegie Mellon University\\
\email{xsh1302@gmail.com}
\and
Haike Xu\\
MIT\\
\email{haikexu@mit.edu}
\and
Alan Zhou\\
Texas A\&M University\\
\email{alan.zhou.0@tamu.edu}
\and
Samson Zhou\\
Texas A\&M University\\
\email{samsonzhou@gmail.com} 
}

\begin{document}

\maketitle

\begin{abstract}
In the $k$-diverse nearest neighbor search problem, the input is a dataset in which each point is assigned a color, and given a query $q$, the goal is to retrieve $k$ approximate nearest neighbors with distinct colors. This color diversity metric is essential for information retrieval tasks where results are required to have different categories, e.g., recommending products from different sellers. Previous state-of-the-art solution by Anand et al. [ICML 2025] constructs a diversity-aware graph to meet this requirement. However, they suffer from an $\mathcal{O}(k^2)$  multiplicative query time overhead, creating a computational bottleneck when $k$ is large.

To break this barrier, we propose a novel color bucketing framework that selects a subset of colors in each bucket and instantiates an independent approximate nearest neighbor search algorithm on the corresponding points. Inspired by group testing, we provide a randomized color sampling scheme that achieves an $\mathcal{O}\left(k^{1-\frac{1}{c^2}-o(1)}\log k\right)$ overhead in Euclidean spaces and an $\mathcal{O}\left(k \log k\right)$ overhead in general metric spaces, where $c$ is the approximation factor, reducing a factor of $k$. In addition, the space usage to store the data structure matches or improves upon prior methods. Empirical results on semi-synthetic and real-world datasets demonstrate that our approach achieves significantly faster search times and higher recall compared to state-of-the-art diversity search baselines.

Furthermore, we introduce a dataset-oblivious deterministic bucketing scheme using expander graphs for a relaxed diverse search problem that only requires $k(1-\varepsilon)$ colors. Here, dataset-oblivious means that the color bucket construction does not depend on the spatial configuration of the dataset. We then establish an $\Omega(k^2)$ lower bound on the number of buckets for any dataset-oblivious deterministic bucketing scheme that guarantees $k$ points of distinct colors within $c$ times the optimal radius. We subsequently circumvent this by a dataset-aware adaptive segment tree bucketing scheme with an overhead of only $\mathcal{O}(k \log N)$, where $N$ is the number of colors. As an extension of our results, we solve the problem with general diversity metric, where the goal is to maximize the minimum pair-wise distance of the $k$ solutions, and improve the query time.
\end{abstract}

\section{Introduction}

In the nearest neighbor search (NNS) problem, the input is a set $P$ of $n$ points in a metric space $\calX$, and the goal is to construct a data structure that, for any query $q$, reports a point $p \in P$ with minimum distance to $q$.
This problem is a fundamental data structure problem with wide-ranging applications in modern computational tasks.
In recommendation systems, NNS is a cornerstone for collaborative filtering and content-based methods, which are significant in identifying the most relevant answer for a user's query.
In computer vision, NNS facilitates large-scale image retrieval and facial recognition by matching feature vectors extracted from deep neural networks to massive databases of known descriptors \cite{jegou2010aggregating}.
Furthermore, in the recent rise of generative AI and Large Language Models (LLMs), NNS serves as an important primitive to retrieve contextually relevant documents from vector databases \cite{MY18,khandelwal2019generalization} and has been used to accelerate attention computation in Transformer-based architectures \cite{han2023hyperattention}.
Given these applications, we have seen a surge of interest in developing efficient algorithms for NNS \cite{IndykM98,SDI06,AndoniI08,MY18,IM18,JSDS+19,IndykX23,AnandIKMRSX25}.

Given that existing data structures for \emph{exact} nearest neighbor search are inefficient for high-dimensional point sets, the following relaxed $c$-approximate nearest neighbor ($c$-ANN) problem is widely adopted, where we allow to report a $c$-approximation to the exact nearest neighbor.

\begin{definition}[The $c$-ANN problem]
\deflab{def:NN}
Given a set $P$ of $n$ points in a metric space $(\calX, \mathrm{dist})$, the goal is to construct a data structure that satisfies the following: given any query $q$, if the nearest neighbor of $q$ is within distance $r$, it reports a point $\widehat{p}$ that satisfies $\mathrm{dist}(\widehat{p},q) \le cr$.
\end{definition}

Classical algorithms address the $c$-ANN problem with efficient space usage and update time, including those by \cite{IndykM98,AndoniI08} using locality sensitive hashing (LSH) in Euclidean spaces, and \cite{KrauthgamerL04} using navigating nets in general metric spaces.
More recently, graph-based algorithms such as HNSW \cite{MY18}, NGT \cite{IM18}, and DiskANN \cite{JSDS+19} have gained increasing attention, as they achieve state-of-the-art performance in practice \cite{ABF24}, provable guarantees \cite{IndykX23}, and apply to general metrics. 
These methods build a navigation graph over the dataset $P$, and the query procedure performs a greedy search on this graph to locate points close to the query.

In many real-world applications, it is often necessary to return not just a single nearest neighbor but several of them, for instance, a search engine typically provides multiple websites in response to a user’s query.
While minimizing the distance of the reported points is the primary goal, it has been observed that the returned results can be redundant or nearly identical \cite{CG98}.
This issue is especially pronounced in applications in recommendation systems, where many similar versions of the same product, seller, or document may exist, e.g., a user buying shoes may prefer to be shown options from different sellers. 
This motivates the goal of diversity-aware search, as emphasized by Google’s 2019 update, which ensures that no more than two pages from the same site are shown \cite{Lia19}.
To accurately capture this requirement, we assign each point a \emph{color}, representing a categorical label such as a site ID or a seller \cite{AnandIKMRSX25}, and require the algorithm to return a set $S$ of $k$ points with distinct colors, while keeping their distances to the query $q$ near-optimal.
We next provide the formal definition of the $k$-diverse $c$-ANN problem.



\begin{definition}[The $k$-diverse $c$-ANN problem]
\deflab{def:k:div:NN}
Given a set $P$ of $n$ points in a metric space $(\calX, \mathrm{dist})$, where each point is assigned a color, the goal is to construct a data structure that satisfies the following: for any query $q$, if the smallest ball around $q$ which contains $k$ points of distinct colors has radius at most $r$, it returns a set $\widehat{S}$ of $k$ points of distinct colors that satisfies $\max_{p \in \widehat{S}}\mathrm{dist}(p,q) \le cr$. A randomized data structure for this problems succeeds with probability $1-\delta$ if, for any fixed $q$, the preceding condition holds except with probability at most $\delta$.
\end{definition}

To solve the $k$-diverse $c$-ANN problem, the early work of \cite{CG98} introduced a two-stage diversity-based re-ranking approach: they first select $m$ closest points to the query, and then post-process the $m$ points to identify $k$ diverse points.
Although widely used, this pipeline has a fundamental efficiency drawback. In practice, the algorithm must retrieve $m \gg k$ points (with $m = \Omega(n)$ in the worst case) to ensure that we recover the $k$ diverse points.
To address this issue, subsequent work modified classical $c$-ANN algorithms to return $k$ diverse points, including the algorithm in \cite{AbbarAIMV13}, which stores $k$ diverse points in each LSH bucket, and the algorithm in \cite{AnandIKMRSX25}, which constructs disk-ANN graphs subject to diversity constraints. 
These results bypass the dependence on $m$, demonstrating that diversity-aware search can be achieved without first retrieving all relevant points. 
However, their query time still scales as $k^2$, which becomes inefficient when $k$ is large, e.g., in recommending hundreds of diverse products to a user. 
This raises a natural question:
\begin{center}
\emph{Is it possible to bypass the quadratic dependence on $k$ in the query time for the $k$-diverse $c$-ANN problem?}
\end{center}
In this work, we provide algorithms that significantly improve on the $k$ factors, breaking the $k^2$ barrier and achieving efficient query time.
The overheads are sublinear in $k$ in Euclidean spaces and linear in $k$ in general metric spaces, with space comparable to or better than previous work.
Therefore, our approaches support applications to large-scale recommendation and retrieval systems. 
Importantly, we provide a unified color bucketing framework that black-boxes standard $c$-ANN algorithms to solve diverse search problems with efficient computational overhead.

\subsection{Our Results}

In this section, we introduce the guarantees of our $k$-diverse $c$-ANN algorithms.  
We present the space and query time by the computational overhead compared to classical $c$-ANN algorithms, for instance, the LSH-based $c$-ANN algorithm in Euclidean spaces uses $\O{dn^{\frac{1}{c^2}+o(1)}}$ time, while our algorithm for $k$-diverse $c$-ANN in Euclidean spaces uses $\O{k^{1-\frac{1}{c^2}-o(1)}\log k \cdot dn^{\frac{1}{c^2}+o(1)}}$ time, then the overhead is $\O{k^{1-\frac{1}{c^2}-o(1)}\log k}$.
This notion allows a fair comparison with previous work \cite{AbbarAIM13,AbbarAIMV13,AnandIKMRSX25}, since those algorithms rely heavily on the data structures in classical $c$-ANN algorithms while our algorithms directly apply $c$-ANN algorithms as black boxes.  
We first present the result of our randomized algorithm, which achieves a query time sublinear in $k$.

\begin{theorem}[Informal statement of \thmref{thm:k:div:euclidean:front}]
\thmlab{thm:rand}
Given a set $P$ of $n$ points in a metric space $(\calX, \mathrm{dist})$, there exists a randomized algorithm that solves the $k$-diverse $c$-ANN problem (see \defref{def:k:div:NN}) with probability at least $0.99$.  
The algorithm achieves the following overheads:  
\begin{itemize}
\item 
$\O{k^{1-1/c^2-o(1)} \log k}$ in query time and $\O{\log k}$ in space in Euclidean spaces;  
\item 
$\O{k \log k}$ in query time and $\O{\log k}$ in space in general metric spaces.  
\end{itemize}
\end{theorem}

For comparison, the algorithm of \cite{AbbarAIMV13} incurs an $\O{k \log k}$ overhead (implicitly) in Euclidean spaces, while our algorithm only has $\O{k^{1-1/c^2-o(1)}\log k}$, achieving sublinear dependence on $k$.
In general metric spaces, the algorithm of \cite{AnandIKMRSX25} incurs an $\O{k^2}$ overhead, while our algorithm only has $\O{k\log k}$, reducing a $k$ factor and bypassing the quadratic barrier.
Furthermore, the $\log k$ factor can be replaced by $\log \tfrac{1}{\varepsilon}$ for the relaxed $(k,\eps)$-diverse $c$-ANN problem, where we allow the output to contain only $\lceil(1-\varepsilon)k\rceil$ points with distinct colors.
Our results are based on a simple and unified color bucketing framework, which selects a subset of colors in each bucket and implements an independent $c$-ANN algorithm on the corresponding points.

We also provide a deterministic algorithm via expander graphs in general metric spaces, achieving improved dependence on $k$ for the relaxed $(k,\eps)$-diverse $c$-ANN problem.
\begin{theorem}[Informal statement of \thmref{thm:k:div:det:front}]
\thmlab{thm:det}
Given a set $P$ of $n$ points in a metric space $(\calX, \mathrm{dist})$, there is a deterministic algorithm that solves the $(k,\eps)$-diverse $c$-ANN problem with an $\O{k^{1+o(1)}\gamma}$ overhead in query time and an $\O{\gamma}$ in space, where $\gamma = \left(\frac{\log k \log N}{\eps}\right)^{\O{\sqrt{\log k}}}$.
\end{theorem}
\vspace{-0.2cm}
We remark that the color bucketing scheme in \thmref{thm:det} is dataset-oblivious, meaning that the color assignment only depends on the color set and problem parameters, and is oblivious to the spatial configuration of the dataset and the query point. 
Requiring all $k$ distinct colors imposes a stronger condition on such a scheme. 
If every bucket containing a color $x$ also contains one of $k-1$ competing colors, we can place those competitors close enough to the query to mask $x$ in every such bucket, even when $x$ is needed for a valid answer. 
A deterministic, dataset-oblivious scheme must therefore isolate each color from every set of $k-1$ other colors. 
This is precisely the $(k-1)$-disjunctness property, which gives the following lower bound on the number of buckets. 

\begin{theorem}[Informal statement of \thmref{thm:lb:front}]
Any dataset-oblivious deterministic bucketing scheme that solves the $k$-diverse $c$-ANN problem (see \defref{def:k:div:NN}) with $N=\Omega(k^2)$ colors must evaluate $L=\Omega\left(k^2 \frac{\log N}{\log k}\right)$ buckets.
\end{theorem} 
\vspace{-0.2cm}
To circumvent the quadratic barrier, we design a dataset-aware deterministic scheme that queries the buckets adaptively based on the dataset and the query, achieving an $\O{k \log N}$ overhead.
\vspace{+0.05cm}

\begin{theorem}[Informal statement of \thmref{thm:det:adap:front}]
\thmlab{thm:det:adap:front'}
Given a set $P$ of $n$ points in a metric space $(\calX, \mathrm{dist})$, there is an adaptive deterministic algorithm that solves the $k$-diverse $c$-ANN problem (see \defref{def:k:div:NN}) with an
$\O{k\log N}$ overhead in query time and an $\O{\log N}$ overhead in space in general metric spaces.
\end{theorem}
\vspace{-0.1cm}
We summarize our main results for the $k$-diverse problem in \tableref{tab:results}. 
We improve on the dependence on $k$ in query time while maintaining comparable or better space compared to prior work.

\begin{table*}[!htb]
\centering
{
\tabulinesep=1.1mm
\begin{tabu}{|c|c|c|c|c|c|}\hline
{\bf References} &{\bf Metric}  & {\bf Query time} & {\bf Space} & {\bf Remark}  \\\hline\hline
\cite{AbbarAIMV13} & Euclidean & $\tO{k}$ & $\tO{1}$ & Randomized  \\\hline
\thmref{thm:rand} & Euclidean & $\tO{k^{1-1/c^2-o(1)}}$ & $\tO{1}$ & Randomized  \\\hline \hline
\cite{AnandIKMRSX25} & General & $\O{k^2}$ & $\O{k}$ & Deterministic  \\\hline
\thmref{thm:rand} & General & $\tO{k}$ & $\tO{1}$ & Randomized  \\\hline
\thmref{thm:det:adap:front'} & General & $\O{k \log N}$ & $\O{\log N}$ & Deterministic  \\\hline
\end{tabu}
}
\caption{Comparison with previous work. 
We use $\tO{\cdot}$ to hide the $\log k$ factor.}
\tablelab{tab:results}
\vspace{-0.5cm}
\end{table*}







\subsection{Technical Novelties}
\label{sec:technical_contributions}
The primary challenge in diverse nearest neighbor search is satisfying categorical constraints without incurring prohibitive search costs. 
While early post-processing approaches \cite{CG98} completely decoupled search from diversity, they required an exhaustive scan of up to $\Omega(n)$ points to guarantee $k$ diverse results. 
Recent methods \cite{AbbarAIM13, AbbarAIMV13, AnandIKMRSX25} instead embed diversity constraints directly into the search index, but this approach forces the query procedure to redundantly traverse dense, heavily overlapping spatial neighborhoods to discover new colors, creating a $\O{k^2}$ query time bottleneck. 
To systemically overcome this quadratic bottleneck, we decouple the spatial search from the diversity constraints, thereby establishing novel connections between diverse retrieval, graph theory, group testing, and coding theory. 

\textbf{Color bucketing framework via group testing.}
We first show that querying for $k$ diverse points can be viewed as a geometric analogue of non-adaptive group testing~\cite{dorfman1943detection,IndykNR10,du1999combinatorial}, where one seeks to identify a small set of ``defective'' items through parallel aggregate tests. 
By treating the $k$ closest uniquely colored points as the defectives and a standard black-box $c$-ANN oracle as the group test, we introduce a randomized color bucketing scheme. 
We prove that assigning colors to $\O{k\log k}$ buckets independently with probability $\frac{1}{k}$ isolates target colors from spatially closer but redundant points with high probability, thereby bypassing the previous $\O{k^2}$ barrier. 

\textbf{Information-theoretic lower bounds for exact search.} 
To understand the fundamental limits of exact deterministic retrieval, we formalize the study of \emph{dataset-oblivious} bucketing schemes, where color assignments are fixed independently of the dataset. 
We identify a geometric ``masking'' phenomenon, where target colors are consistently occluded by closer, redundant points across queries. 
We prove a strict equivalence between avoiding this masking and algebraic coding theory: for any static scheme to guarantee exact recovery, its color-to-bucket incidence matrix must form a \emph{separable matrix}, i.e., a $(k-1)$-disjunct matrix or $(1, k-1)$-cover-free family, meaning no column is contained in the union of any $k-1$ others. 
By translating spatial diversity into this combinatorial requirement, we leverage bounds on disjunctive codes from extremal combinatorics to establish an $\Omega\left(k^2 \frac{\log N}{\log k}\right)$ lower bound on the index size, thus proving that the quadratic overhead observed in prior deterministic algorithms is an inherent information-theoretic barrier. 

\textbf{Bypassing bounds via expander graphs.} 
Motivated by this impossibility result, we consider the $(k,\eps)$-diverse problem and design a dataset-oblivious scheme based on unbalanced bipartite expander graphs. 
We show that the vertex expansion properties of these graphs preserve spatial metric neighborhoods sufficiently well to guarantee the isolation of most target colors, without strictly requiring the rigid $(k-1)$-disjunctness property. 
This relaxation allows us to safely drop the deterministic query overhead to $\O{k^{1+o(1)}}$.

\textbf{Adaptive segment trees for deterministic $k$-diverse search.} 
To achieve \emph{exact} $k$-diverse recovery without the coding-theoretic penalty, we shift to a dynamic, dataset-aware approach. 
We construct an adaptive segment tree over the color space that interleaves categorical partitioning with spatial priority search. 
During a query, the algorithm dynamically routes down the tree, computing mutually disjoint partitions of the remaining active colors conditioned on the geometry of candidates retrieved on the fly. 
This adaptive routing sidesteps the worst-case covering bounds of static matrices entirely, achieving exact deterministic recovery with only an $\O{k \log N}$ overhead.

\textbf{Sub-quadratic thresholding for continuous diversity.} 
We extend our framework to spatial diversity objectives maximizing minimum pairwise distance. 
Prior LSH-based pipelines \cite{AbbarAIM13} rely on Greedy Maximal Margin selection, incurring $\O{mk}$ time from exhaustive distance checks. 
We replace this with an LSH-augmented thresholding scheme that selects points via approximate nearest-neighbor queries while maintaining a dynamic index, eliminating explicit pairwise comparisons. 
This gives a constant-factor approximation in sub-quadratic time, reducing overhead to $\O{mk^{1/c^2}}$.

\subsection{Notations}

We introduce the notations used in this paper.
We refer the readers to \secref{sec:prev:ann} for a preliminary introduction on standard $c$-ANN algorithms.
Let $P$ be a set of $n$ points in a metric space $(\calX, \mathrm{dist})$, e.g., the Euclidean space $(\mathbb{R}^d, \ell_2)$ and the Hamming metric space $(\{0,1\}^d, \ell_1)$. 
Let $\dist(x,S) = \min_{s \in S} \dist(x,s)$ be the distance between a point $x \in P$ and a set $S \subset P$.
Let $\calC$ be the set of colors with size $N$, and for a point $p \in P$, let $c(p)$ be its color.
Let $\Delta$ denote the aspect ratio, which is the ratio between the diameter of $P$ and the minimum pair-wise distance of $P$, i.e., $\min_{p,p' \in P} \dist(p,p')$. 
Let $B(p,r)= \{x \in \calX, \dist(x,p)<r\}$ be the ball of radius $r$ centered at $p$.
Let $\mathbf{ddim}$ denote the doubling dimension of $P$, which is the smallest integer such that for any ball of radius $r$ centered at some point $p \in P$, the set $B(p,r) \cap P$ can be covered by at most $2^{\mathbf{ddim}}$ balls of radius $r/2$. 



\section{Randomized Algorithms for Diverse Nearest Neighbor Search}
\seclab{sec:color:sampling}

In this section, we introduce our color sampling approach inspired by group testing.
In group testing, the goal is to identify a small set of “defective” items by grouping together subsets of items and testing them collectively, rather than testing each one individually. 
A positive test guarantees that at least one defective item lies in the group, while a negative test eliminates the entire group. 
Suppose that there are $k$ defectives, a randomized algorithm for group testing is by sampling an item with probability $\frac{1}{k}$ in each test, and this identifies the $k$ defectives using $\O{k \log k}$ tests.

In our setting, the defectives correspond to the $k$ distinct colors of the target set, i.e., the set of $k$ points with distinct colors and has closest distance to the query $q$. 
Based on this observation, in each bucket, we sample each color with probability $\frac{1}{k}$ and implement a separate $c$-ANN algorithm.
The guarantees of the $c$-ANN algorithm and group testing then ensure that $\O{k \log k}$ randomized buckets collectively select a set of nearest neighbors with distinct colors, satisfying the requirement of $k$-diverse $c$-ANN problem.
We now formalize this approach in \algref{alg:div:group:testing}.
We summarize the guarantees of our algorithm in \thmref{thm:k:div:euclidean:front} and defer the formal analysis to \secref{sec:color:sampling:proof}. 
We note that our algorithm achieves sublinear overhead in $k$ in Euclidean spaces and Hamming spaces and near-linear overhead in general metric spaces.

\seclab{sec:div:rand}
\begin{algorithm}[!ht]
\caption{Diversity search by color sampling}
\alglab{alg:div:group:testing}
\begin{algorithmic}[1]
\Require{Input dataset $P = \{p_1, \cdots, p_n\}$, each with a color $c \in \calC$, an integer $k \le |\calC|$}
\Ensure{Data structure to solve the $k$-diverse nearest neighbor problem}
\State{Let $L \gets \O{k \log k}$}
\For{$i \in [L]$}
\State{Sample each color $c \in \calC$ with probability $\frac{1}{k}$}
\State{Let $\calC_i$ be the set of sampled colors, and let $P_i$ be the set of points with colors in $\calC_i$}
\State{Maintain a data structure that solves the $c$-ANN problem for $P_i$} 
\EndFor
\For{each query $q$}
\State{Let $\widehat{p_i}$ be the reported $c$-ANN of bucket $i$ for each $i \in [L]$}
\State{{\bf Return:} $k$ nearest points to $q$ in $\{\widehat{p_1} \ldots, \widehat{p_L}\}$ of distinct colors}
\EndFor
\end{algorithmic}
\end{algorithm}

\begin{theorem}
\thmlab{thm:k:div:euclidean:front}
Given a set $P$ of $n$ points, there exist algorithms that solve the $k$-diverse $c$-ANN problem (see \defref{def:k:div:NN}) with probability $0.99$ in the following spaces:
\begin{itemize}
\item 
Euclidean spaces $(\mathbb{R}^d, \ell_2)$: the algorithm uses query time $\O{k^{1-\frac{1}{c^2}-o(1)}\log k \cdot dn^{\frac{1}{c^2}+o(1)}}$ and space $\O{\log k \cdot dn^{1+\frac{1}{c^2}+o(1)}}$.
\item 
Hamming spaces $(\{0,1\}^d, \ell_1)$: the algorithm uses query time $\O{k^{1-\frac{1}{c}-o(1)}\log k \cdot dn^{\frac{1}{c}+o(1)}}$ and space $\O{\log k \cdot dn^{1+\frac{1}{c}+o(1)}}$. 
\item 
General metric spaces $(\calX, \dist)$: the algorithm uses  $\O{k\log k \cdot (4\alpha)^{\mathbf{ddim}} \log_\alpha \frac{\Delta}{(\alpha-1)\eps} \log \Delta}$ query time and $\O{\log k\cdot n(4\alpha)^{\mathbf{ddim}} \log \Delta}$ space, where $\mathbf{ddim}$ is the doubling dimension and $\Delta$ is the aspect ratio.
\end{itemize}
\end{theorem}

We extend our method to solve two variants of the $k$-diverse $c$-ANN problem.
First, using only $L=\O{k \log \frac{1}{\eps}}$ buckets in our algorithm solves the relaxed $(k,\eps)$-diverse $c$-ANN problem (see \defref{def:k:eps:div:NN}), improving the $\log k$ factor in the overhead to a $\log \frac{1}{\eps}$ factor (see \secref{sec:color:sampling:proof}).
Second, sampling each color with probability $\frac{k'}{2k}$ and return $k'$ $c$-ANNs, instead of one, in each bucket solves the $(k,k')$-diverse $c$-ANN problem (see \defref{def:k':nn}), which incurs an $\O{k^{1-\frac{1}{c^2}-o(1)}(k')^{\frac{1}{c^2}+o(1)}\log k}$ overhead in query time, improving the overheads obtained by \cite{AnandIKMRSX25} (see \secref{sec:red}).
These results demonstrate the adaptivity of our method.

\vspace{-0.1cm}
\section{Deterministic Algorithms for Diverse Nearest Neighbor Search}

In this section, we provide the construction and lower bound for deterministic algorithms for the $k$-diverse $c$-ANN problem.
To obtain deterministic algorithms, we replace random sampling with deterministic bucketing and querying constructions. 

\subsection{Graph Expander Construction}
\seclab{sec:expander}

First, we provide our dataset-oblivious color bucketing scheme via graph-expander.
Expanders are bipartite graphs with strong neighborhood expansion properties: any small subset of left vertices has many right neighbors. 
We next formalize the definition of a $(k,\Gamma)$-expander.

\seclab{sec:div:det}
\begin{definition}[Expander]
Let $G=(A,B,E)$ be a bipartite graph, where $A$ is the set of left vertices with size $N$ and left-degree $D$, $B$ is the set of right vertices with size $M$, and $E: [N] \times [D] \to [M]$ is the edge list.
Let $k \le |A|$ and $\Gamma$ be two integers.
$G$ is a $(k,\Gamma)$-expander if for every set $S \subset A$ such that $|S| \le k$, $|N(S)| \ge \Gamma \cdot |S|$, where $N(S)$ is the neighborhood of $S$.
\end{definition}
\vspace{-0.2cm}
The following theorem shows a deterministic construction of expanders with small degree introduced by \cite{GuruswamiUV09}, which ensures the efficiency of our algorithm. 
\begin{theorem}[\cite{GuruswamiUV09}]
\thmlab{thm:expander}
For all positive integers $k,N$, all $\eps > 0$, and all $\gamma \in
(0, \frac{\log x}{\log \log x})$ for $x=\frac{\log k \log N}{\eps}$, there is an explicit $(k,(1-\eps)D)$-expander $G=(A,B,E)$ with $E: [N] \times [D] \to [M]$ such that $D = \O{\left(\frac{\log k \log N}{\eps}\right)^{1+\frac{1}{\gamma}}}$ and $M \le D^2 k^{1+\gamma}$.
\end{theorem}
\vspace{-0.3cm}
In our setting, the left vertices in $A$ represent the colors, and the right vertices in $B$ represent the buckets, and each bucket includes the points with colors connected to it in the expander.
The expander property in \thmref{thm:expander} guarantees that for every set of $k$ colors, there exist at least $k(1-\eps)D$ buckets in their neighbor set, ensuring that at least $k(1-\eps)$ colors are found since each color has degree $D$.
In addition, we query at most $M \le D^2 k^{1+\gamma}$ buckets, giving an overhead of roughly $\O{k^{1+o(1)}}$.
We note that the DiskANN algorithm with provable guarantee in \cite{IndykX23} is deterministic, so combining it with our expander-based color bucketing approach gives a deterministic algorithm.
We defer the formal discussion of the above process to \algref{alg:div:expander:back} in \secref{sec:expander:proof}.

\begin{theorem}
\thmlab{thm:k:div:det:front}
Given a set $P$ of $n$ points in a metric space $(\calX, \mathrm{dist})$ with doubling dimension $\mathbf{ddim}$, aspect ratio $\Delta$, and $N$ colors, there is an algorithm that solves the $(k,\eps)$-diverse $c$-ANN problem using space $\O{\gamma \cdot n(4\alpha)^{\mathbf{ddim}} \log \Delta}$, where $\gamma = \left(\frac{\log k \log N}{\eps}\right)^{\O{\sqrt{\log k}}}$, and $\O{k^{1+o(1)}\gamma\cdot(4\alpha)^{\mathbf{ddim}} \log_\alpha \frac{\Delta}{(\alpha-1)\eps} \log \Delta}$ query time.
\end{theorem}
\vspace{-0.2cm}

\subsection{Lower Bound for Dataset-Oblivious Bucketing}
\seclab{sec:lb:front}

In this section, we show a lower bound on deterministic dataset-oblivious color bucketing schemes.
The key observation is that for dataset-oblivious scheme that solves the $k$-diverse problem, the color assignment must satisfy $k$-disjunctness.
If not, there exists at least a color $x^* \in \mathcal{C}$ and a subset of colors $S^* \subset \mathcal{C} \backslash\left\{x^*\right\}$ of size $k-1$, such that every bucket containing $x^*$ also contains at least one color from $S^*$.
Because the scheme is dataset-oblivious, it must be universally correct for any spatial configuration of the dataset.
Then consider dataset $P$ and query $q$ satisfying the following properties:
\begin{enumerate}
\item  For every color $\gamma \in \mathcal{C}$, there is exactly one point $p_\gamma \in P$ with color $\gamma$. 
\item For all $y \in S^*, \operatorname{dist}\left(q, p_y\right)=1$.
\item $\operatorname{dist}\left(q, p_{x^*}\right)=c+\eps$, for an arbitrarily small $\eps>0$.
\item Let $K^*=S^* \cup\left\{x^*\right\}$. For all $z \in \mathcal{C} \backslash K^*, \operatorname{dist}\left(q, p_z\right)=c(c+\eps)+\eps$.
\end{enumerate}

In the above construction, any point outside $K^*$ is not a $c$-ANN, so we must return exactly $K^*$ in our solution.
However, for any bucket that contains $x^*$, another color $y \in S^*$ also exists, then the $c$-ANN algorithm would not report $x^*$ since it is not a $c$-ANN in that bucket due to $\dist(q,p_y)=1$, which leads to a contradiction since $x^*$ must be reported.
Then by information-theoretic lower bounds for $k$-disjunctness, we have the following lower bound.
We defer the formal analysis to \secref{sec:lb}. 
\begin{theorem}
\thmlab{thm:lb:front}
Assuming $k \ge 3$ and $k^2 \frac{\log N}{\log k} \le C \cdot N$ for some constant $C$, any deterministic dataset-oblivious bucketing scheme that correctly solves the $k$-diverse $c$-ANN problem for all valid datasets must evaluate $L=\Omega\left(k^2 \frac{\log N}{\log k}\right)$ buckets.
\end{theorem} 

\subsection{Adaptive Segment Tree Bucketing}
\seclab{sec:adap:front}
To circumvent the $\Omega(k^{2})$ barrier, we introduce an adaptive deterministic algorithm that selectively queries buckets based on the specific spatial configuration and color assignment of the dataset, using a segment tree with $N$ leaves over the color space. 
Unlike the relaxed guarantee in \secref{sec:expander}, the segment tree returns $k$ points of pairwise distinct colors, while retaining the same $c$-approximation to the optimal radius. 
To construct the tree, we assign a different color to each of the $N$ leaves, and for each node $u$ in $\calT$, we include all colors associated with the leaves of the subtree rooted at $u$.
Then a separate $c$-ANN algorithm is ran on each node.

When receiving a query $q$, the algorithm dynamically maintains a priority queue $Q$ that stores a list of pairs $(\widehat{p_u},u)$ in the order of their distance to $q$.
Here, $(\widehat{p_u},u)$ denotes a candidate point $\widehat{p_u}$ and the node $u$ that reports $\widehat{p_u}$. 
We initialize the queue by querying the root node.
Next, in each iteration, we first add the minimum priority element $(\widehat{p_u},u)$ in $Q$ to the result set, then query a group of descendent nodes of $u$ to add new candidates to $Q$ such that we avoid $c(p_u)$ while ensuring that all other colors in $u$ are considered.
To achieve these requirements, we look at the downward path $v_{0}=u,v_{1},...,v_{m}$ in $\mathcal{T}$ from $u$ to the leaf corresponding to $c(\widehat{p_u})$.
Note that for each $i \in [m]$, $v_i$ contains $c(\widehat{p_u})$, so we instead query the other child $w_{i-1}$ of $v_{i-1}$ that is not $v_{i}$ and add its reported point to $Q$.
The key observation is that $\{w_i \mid i\in [m-1] \}$ forms a mutually disjoint partition of the colors in $u$ except for $c(\widehat{p_u})$.
Therefore, we maintain an invariance in each iteration that the remaining nodes in $Q$ forms a mutually disjoint partition of $\calC \setminus \calC_\calR$, where $\calC_\calR$ is the set of colors in the current result set.
This implies that a new valid $c$-ANN with a distinct color is extracted from $Q$ in each iteration, so the algorithm terminates in $k$ iterations.
Since the length of the path is at most $\log N$, we have $\O{k \log N}$ overhead in query time in total.
We formalize the above procedure in \algref{alg:query:det} and defer the formal analysis to \secref{sec:adap}.

\begin{algorithm}[!ht]
\caption{Adaptive query scheme for $k$-diverse $c$-ANN}
\alglab{alg:query:det}
\begin{algorithmic}[1]
\Require Query $q\in\mathcal{X}$, target diversity $k\le N$, segment tree $\mathcal{T}$
\Ensure{Solution to $k$-diverse search}
\State Initialize a priority queue $Q$, ordering pairs $(\widehat{p_u},u)$ by $dist(q,\widehat{p_u})$. Query $\calA_{\text{root}}$ to obtain a $c$-ANN $\widehat{p_{\text{root}}}$ and insert $(\widehat{p_{\text{root}}}, \text{root})$ into $Q$
\State Initialize result set $\mathcal{R}\leftarrow\emptyset$ and active colors $\calC_\calR\leftarrow\emptyset$
\While{$|\mathcal{R}|<k$}
    \State Extract the pair $(\widehat{p_u},u)$ with minimum distance to $q$ from $Q$
    \State $\mathcal{R}\leftarrow\mathcal{R}\cup\{\widehat{p_u}\}$, $\calC_\calR\leftarrow \calC_\calR\cup\{c(\widehat{p_u})\}$
    \State Let $v_{0}=u,v_{1},...,v_{m}$ be the downward path in $\mathcal{T}$ from $u$ to the leaf corresponding to $c(\widehat{p_u})$
    \For{$i=0$ to $m-1$}
        \State Let $w_{i}$ be the child of $v_{i}$ that is not $v_{i+1}$.
        \If{$P_{w_{i}}\ne\emptyset$}
        \State Query $\calA_{w_i}$ to obtain a $c$-ANN $\widehat{p_{w_i}}$ and insert $(\widehat{p_{w_i}}, w_i)$ into $Q$
        \EndIf
    \EndFor
\EndWhile
\Return $\mathcal{R}$
\end{algorithmic}
\end{algorithm}

Combining our algorithm with diskANN \cite{IndykX23}, we have the following result.
\begin{theorem}[$k$-diverse $c$-ANN with general metrics]
\thmlab{thm:det:adap:front}
Given a set $P$ of $n$ points in a metric space $(\calX, \mathrm{dist})$ with doubling dimension $\mathbf{ddim}$ and aspect ratio $\Delta$, there is a deterministic algorithm that solves the $k$-diverse $c$-ANN problem (see \defref{def:k:div:NN}) for $c=\frac{\alpha+1}{\alpha-1}+\eps$ using 
$\O{k\log N \cdot (4\alpha)^{\mathbf{ddim}} \log_\alpha \frac{\Delta}{(\alpha-1)\eps} \log \Delta}$ query time and $\O{\log N\cdot n(4\alpha)^{\mathbf{ddim}} \log \Delta}$ space.
\end{theorem}

\section{Experiments}
In this section, we present experiments on semi-synthetic and real datasets to demonstrate the validity and practicality of our approach. We performed all of our experiments on a Windows Intel i7 desktop with 32 GB of RAM. Empirical results in \cite{AnandIKMRSX25} demonstrate that their diverse search algorithms outperform algorithms that run standard search and perform post-processing on reported candidates. Thus, we use their methods as a baseline for comparison.
Our algorithms achieve faster query time and more accurate answers on both semi-synthetic and real-world datasets, validating the optimality of our color sampling framework.

\paragraph{Datasets.}
We run our algorithms on the following datasets:
\begin{enumerate}
\item 
\emph{ArXiv OpenAI Embeddings} \cite{Emb24}: 
Like \cite{AnandIKMRSX25}, we consider the public real-world ArXiv dataset containing $1536$-dimensional vectors obtained from OpenAI embeddings of paper abstracts. \figref{fig:oai100k} uses the 100K partition  and \figref{fig:oai2m} uses the full 2M partition. Since this dataset does not contain color information, we synthetically assign colors to each vector using the probabilistic distribution proposed by \cite{AnandIKMRSX25}. 
Specifically, for each vector, its color is selected uniformly at random from a primary set $\{1, 2, 3\}$ with probability $0.9$, and is selected uniformly from an expanded set $\{4, \dots, 1000\}$ with probability $0.1$. 
This assignment is motivated by very skewed real-world scenarios, i.e., a dominant seller in a market, that necessitate diversity in search results.
\item 
\emph{LAION} \cite{laion}: Second, we consider the real-world LAION dataset containing $4$ millions image embeddings of $512$ dimensions, where the embeddings are generated using CLIP \cite{RadfordKHRGASAM21}. This dataset has previously been used to examine graph-based nearest neighbor search algorithms for hybrid search \cite{PatelKGZ24}. Each image in the dataset has multiple keywords that describe its content, which are used to define the colors in our experiments. See \figref{fig:laion5m:dist} for the color distribution.
\item 
\emph{ArXiv-for-fanns} \cite{fanns}: 
We also consider the real-world ArXiv dataset containing $4096$-dimensional vectors obtained by embedding abstracts via a Transformer-based model  \cite{stella-1,stella-2}. \figref{fig:fanns100k} uses the 100K partition and \figref{fig:fanns2M} uses the full 2.8M partition. For feasibility, we use a 100K partition in our experiments. Previously, this dataset has been used to benchmark diverse nearest neighbor search algorithms \cite{fanns}. Since this dataset is composed of paper abstract embeddings, each item is associated with a main category, e.g., math, computer science, quantum physics, etc. There are 30 unique categories, which we use as the ``colors.'' See \figref{fig:fanns100k:dist} and \figref{fig:fanns2M:dist} for the color distribution.

\end{enumerate}


\paragraph{Methods.} For the OpenAI Embeddings dataset, we aim to retrieve $k=100$ nearest neighbors with distinct colors. The LAION dataset only contains 28 unique categories, so we aim to retrieve $k=5$ nearest neighbors of distinct colors. Recall that, \cite{AnandIKMRSX25}'s algorithm constructs a diverse navigation graph based on the graph structure of DiskANN \cite{JSDS+19} and achieves state-of-the-art performance in diverse search.
In contrast, our approach randomly groups the colors into buckets and uses the standard DiskANN graph in each bucket.
As such, we demonstrate our algorithm's performance by using \cite{AnandIKMRSX25}'s algorithm as a baseline for comparison.


\paragraph{Evaluation.} 
We follow the same procedure as \cite{JSDS+19} and \cite{AnandIKMRSX25} to measure the performance of our algorithm. For some query $q$, let the true $k$ nearest neighbors of different colors be $X$ and let the inferred neighbors be $\hat{X}$. Then, recall is defined as $\frac{\lvert X\cap\hat{X}\rvert}{k}$. For our algorithm, we vary the recall by changing the bucket count, and, for the baseline, we vary it by changing the search list size. For various recalls, we measure the mean latency in answering 20K queries for the OpenAI Embeddings dataset and 10K queries for the LAION and ArXiv-for-fanns datasets, allowing us to observe the recall-vs-latency tradeoff. We also analyze throughput by measuring queries per second (QPS) for both algorithms.

\begin{figure}[!h]
    \centering
    \begin{subfigure}{0.31\textwidth}
        \centering
        \includegraphics[width=\textwidth, height=4.0cm]{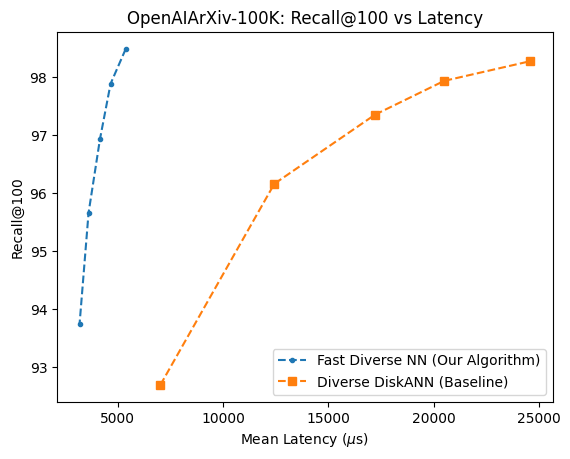}
        \subcaption{Recall-latency tradeoff}
        \figlab{fig:oai100k:lat}
    \end{subfigure}
    \hfill
    \begin{subfigure}{0.31\textwidth}
        \centering
        \includegraphics[width=\textwidth, height=4.0cm]{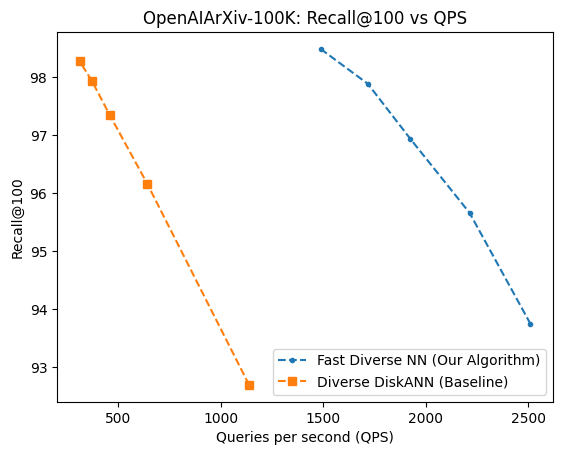}
        \subcaption{Throughput tradeoff}
        \figlab{fig:oai100k:qps}
    \end{subfigure}
    \hfill
    \begin{subfigure}{0.31\textwidth}
        \centering
        \includegraphics[width=\textwidth, height=4.0cm]{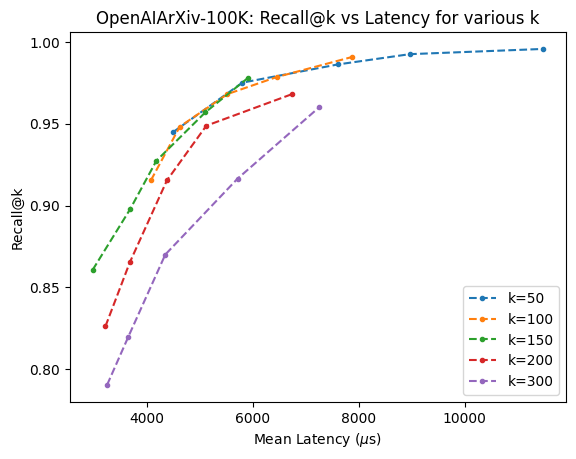}
        \subcaption{Recall-latency tradeoff}
        \figlab{fig:oai100k:abl}
    \end{subfigure}
     \caption{Comparison on OpenAI Embeddings 100K Dataset}
     \figlab{fig:oai100k}
\end{figure}

\begin{figure*}[!h]
    \centering
    \begin{subfigure}{0.31\textwidth}
        \centering
        \includegraphics[width=\textwidth, height=4.0cm]{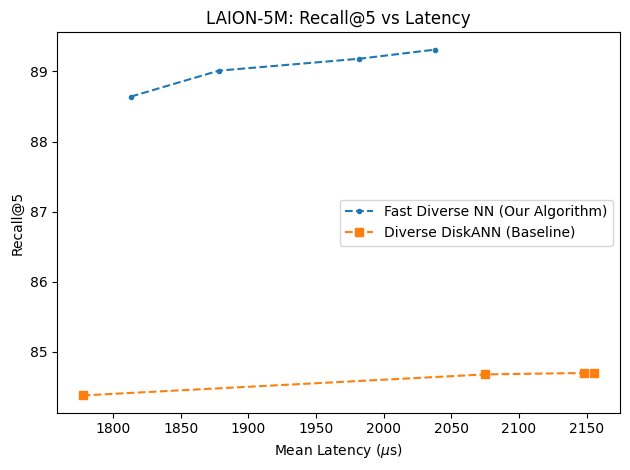}
        \subcaption{Recall-latency tradeoff}
        \figlab{fig:laion5m:lat}
    \end{subfigure}
    \hfill
    \begin{subfigure}{0.31\textwidth}
        \centering
        \includegraphics[width=\textwidth, height=4.0cm]{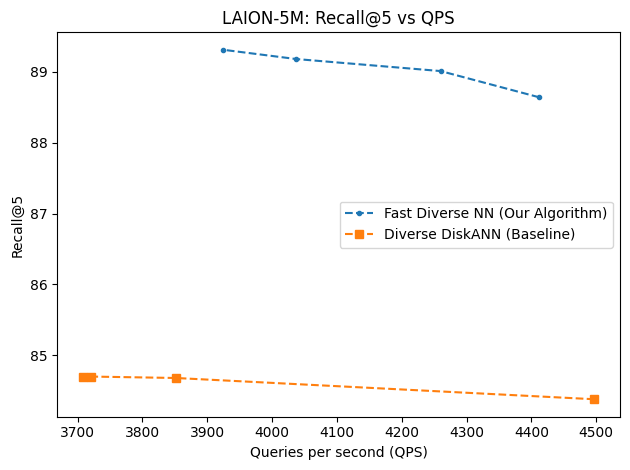}
        \subcaption{Throughput tradeoff}
        \figlab{fig:laion5m:qps}
    \end{subfigure}
    \hfill
    \begin{subfigure}{0.31\textwidth}
        \centering
        \includegraphics[width=\textwidth, height=4.0cm]{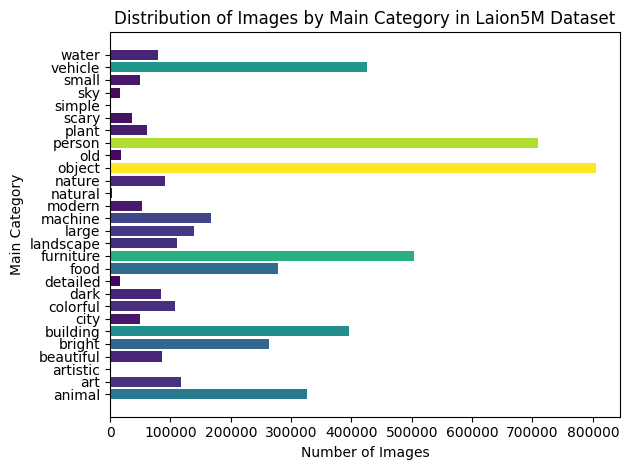}
        \caption{Image category distribution}
        \figlab{fig:laion5m:dist}
    \end{subfigure}
     \caption{Comparison on real-world LAION dataset}
     \figlab{fig:laion5m}
\end{figure*}

\paragraph{Parameter setup.} We configured the baseline algorithm through multiple tests on different graph degree ($R$) and list size ($L$) combinations. For both versions of the OpenAI Embeddings dataset, we used $R=64$ and $L=128$ and varied $L$ during search between 100 and 500. For the ArXiv-for-fanns dataset, we used $R=128$ and $L=128$ and used $L \in \set{3, 9, 27, 81, 243}$ during search. For the LAION dataset, we utilized $R=64$, $L=64$, and used $L\in\set{5, 10, 15, 20, 25}$ during search time. For our algorithm, we used $R=16$ and $L=16$ for the OpenAI and Arxiv-for-Fanns Datasets and $R=64$ and $L=64$ for the LAION dataset. All evaluations used 8 threads during search time. 

\paragraph{Results.}

\figref{fig:oai100k:lat} and \figref{fig:oai100k:qps} compare the recall-latency and recall-QPS tradeoff of our algorithm to the baseline's on the semi-synthetic OpenAI Embeddings dataset. Our algorithm's recall is in a higher range (93.75 - 98.48) than the baseline's recall (92.70 - 98.20). As latency increases, the gap between the algorithms widens, highlighting that our algorithm provides higher recall for the fixed latency. 
\figref{fig:oai100k:abl} shows the recall-latency tradeoff for various $k$ on the OpenAI Embeddings dataset. Note that these experiments utilized different parameters. Specifically, we set $R=64$ and $L=128$, assigned points from the same skewed distribution but with $1987$ unique colors, and set buckets count approximately between $6k$ and $10k$. We notice that across the various values of $k$, our algorithm achieves generally good performance, with recall ranging from $79.05$ to $99.57$. Additionally, the latency is comparable to that in \figref{fig:oai100k}, even with higher graph degree and list size, so the performance can likely be further optimized by increasing the bucket count. 

\figref{fig:laion5m} compares the recall-latency and recall-QPS tradeoff of our algorithm to that of the baseline on the real-word LAION dataset. Our results are consistent with the other datasets, as we have a higher recall range (88.64 - 89.31) than the baseline (80.82 - 83.71). For the same latency and QPS, our algorithm produces higher recall. As latency increases, the gap between the algorithms grows. \figref{fig:laion5m:dist} shows the distribution of main categories or ``colors'' in the experiments involving the LAION dataset. This distribution is considerably less skewed than the synthetically generated one for the OpenAI Embeddings dataset, but our algorithm maintains its advantage, highlighting that it can be deployed in practical tasks like finding $k$ images of distinct objects.

\section{Conclusion}
Our color-bucketing framework reduces $k$-diverse nearest neighbor search to standard $c$-ANN queries. 
In general metric spaces, our randomized algorithm improves the query-time overhead from $\O{k^2}$ to $\O{k\log k}$, with an $\O{\log k}$ space overhead. 
As in prior work, the geometric parameters can still dominate the cost. 
For example, one general-metric implementation incurs a factor of $(4\alpha)^{\mathrm{ddim}}$ and also depends on the aspect ratio $\Delta$. 
Since each bucket uses a $c$-ANN data structure as a black box, improvements in its dependence on either parameter carry over directly.

The main open question is the extent of how much the $\O{k\log k}$ query-time overhead can be improved for randomized approaches on general metrics. 
Our deterministic lower bound relies on a spatial configuration in which $k-1$ closer colors can mask a required color. 
To succeed on every such configuration, a dataset-oblivious bucketing scheme must isolate each color from every set of $k-1$ others. 
This forces its incidence matrix to be $(k-1)$-disjunct and requires $\Omega(k^2\log N/\log k)$ buckets when there are $N=\Omega(k^2)$ colors. 
For each fixed query, however, independent random sampling suffices to recover $k$ distinct colors with probability at least $0.99$ using $\O{k\log k}$ buckets. 
Neither the deterministic argument nor known lower bounds for non-adaptive group testing establish a matching lower bound for randomized approaches. 
It also remains open whether the $\O{\log k}$ space overhead can be reduced while preserving the query-time guarantee.

\section*{Acknowledgments}
The authors are grateful to Piotr Indyk for the insightful discussions during the early stages of this project.
Justin Y. Chen was supported by an NSF Graduate Research Fellowship under Grant No. 17453.
Shenghao Xie is supported in part by NSF CCF-2335411, the work was completed when the author was affiliated with Texas A\&M University.
Samson Zhou is supported in part by NSF CCF-2335411 and gratefully acknowledges funding provided by the Oak Ridge Associated Universities (ORAU) Ralph E. Powe Junior Faculty Enhancement Award.



\bibliography{references}
\bibliographystyle{alpha}

\appendix

\section{Algorithms for Standard Nearest Neighbor Search}
\seclab{sec:prev:ann}
In this section, we summarize existing approaches for the standard $c$-ANN problem, which are applied in our diverse algorithms. 

\paragraph{Euclidean spaces.}
For Euclidean spaces, \cite{IndykM98} introduced the prominent Locality Sensitive Hashing (LSH) technique, which constructs hash functions so that nearby points collide with higher probability than far points.
By storing the dataset in multiple hash tables and probing only a small fraction of them, LSH achieves efficient query time in high dimensions.
We next state the formal definition of LSH.

\begin{definition}[Locality-Sensitive Hashing]
\deflab{def:lsh}
Given the Euclidean space $(\calX, \mathrm{dist})$, a hash family $\mathcal{H} = \{h : \calX \to U\}$ is a $(r_1, r_2, p_1, p_2)$-sensitive hashing if, for any points $p, q \in \calX$, the following conditions hold:
\begin{itemize}
    \item If $\mathrm{dist}(p, q) \leq r_1$, then $\Pr_{\mathcal{H}}[h(p) = h(q)] \geq p_1$,
    \item If $\mathrm{dist}(p, q) \geq r_2$, then $\Pr_{\mathcal{H}}[h(p) = h(q)] \leq p_2$.
\end{itemize}   
\end{definition}

Intuitively, LSH hashes close pairs of points to the same bucket and far pairs of points to different buckets, so that we only need to search through a small amount of buckets that contains the query point.
In the $c$-ANN algorithm \cite{IndykM98}, given a family of hash functions defined as \defref{def:lsh}, we construct $L$ composite hash functions $g_1, g_2, \dots, g_L$, each associated with a hash table $A_1, A_2, \dots, A_L$. 
Each hash function $g_i = \langle h_{i,1}, \dots, h_{i,K} \rangle$ is formed by selecting $K$ functions $h_{i,j}$ uniformly at random from $\mathcal{H}$ and can be computed in $\O{dn^{o(1)}}$ time. 
For a data point $p \in \calX$, it is stored in bucket $g_i(p)$ of table $A_i$ for each $i \in \{1, \dots, L\}$. 
To process a query point $q$, the algorithm retrieves points from the union of buckets $A_1(g_1(q)) \cup \dots \cup A_L(g_L(q))$ and selects the approximated nearest neighbor, i.e., only the hash bucket corresponding to $q$ in each table is examined.
We obtain the results with the Hamming metric (see \thmref{thm:cnn:hamming}) using $L=n^{\frac{1}{c}+o(1)}$ hash functions and the results in Euclidean spaces (see \thmref{thm:cnn:euclidean}) using $L=n^{\frac{1}{c^2}+o(1)}$ hash functions.
The next statement shows the result given by \cite{IndykM98} with the Hamming metric.

\begin{theorem}[$c$-ANN with the Hamming metric, \cite{IndykM98}]
\thmlab{thm:cnn:hamming}
Given a set $P$ of $n$ points in the Hamming metric space $(\{0,1\}^d, \ell_1)$, there exists an algorithm that solves the $c$-ANN problem with probability $0.99$, using $\O{dn^{\frac{1}{c}+o(1)}}$ query time and $\O{dn^{1+\frac{1}{c}+o(1)}}$ space and preprocessing.
\end{theorem}

Based on the previous work, \cite{AndoniI08} designed a family of LSH for the Euclidean metric, achieving efficient algorithms in the Euclidean space as shown below.

\begin{theorem}[$c$-ANN in Euclidean spaces, \cite{AndoniI08}]
\thmlab{thm:cnn:euclidean}
Given a set $P$ of $n$ points in the Euclidean space $(\mathbb{R}^d, \ell_2)$, there exists an algorithm that solves the $c$-ANN problem with probability $0.99$, using $\O{dn^{\frac{1}{c^2}+o(1)}}$ query time and $\O{dn^{1+\frac{1}{c^2}+o(1)}}$ space and preprocessing.
\end{theorem}

\paragraph{General metric spaces.}
For general metric spaces, LSH does not apply. 
Instead, \cite{KrauthgamerL04} introduced the navigating nets, which recursively partition the metric space into a hierarchy of nets at geometrically decreasing scales, enabling efficient search by “navigating” through progressively finer scales.
The next statement shows the bounds on the space and query time of the $c$-ANN algorithm by the doubling dimension.
\begin{theorem}[$c$-ANN with general metrics, \cite{KrauthgamerL04}]
\thmlab{thm:cnn:metric:old}
Given a set $P$ of $n$ points in a metric space $(\calX, \mathrm{dist})$ with doubling dimension $\mathbf{ddim}$ and aspect ratio $\Delta$, there exists an algorithm that solves the $c$-ANN problem, using $\O{2^{\O{\mathbf{ddim}}} \log \Delta + \left(\frac{1}{c-1}\right)^{\O{\mathbf{ddim}}}}$ query time and $\O{n2^{\O{\mathbf{ddim}}}}$ space.
\end{theorem}

Recently, \cite{IM18, MY18}, and \cite{IndykX23} studied graph-based algorithms, which achieve remarkable results in practice.
These algorithms build a graph where vertices correspond to data points and edges connect nearby points. 
At query time, the algorithm performs a greedy search in the edge-list of an entry point based on the distance to the query.
The following theorem shows the guarantee of the diskANN algorithm.
\begin{theorem}[$c$-ANN with general metrics, \cite{IndykX23}]
\thmlab{thm:cnn:metric}
Given a set $P$ of $n$ points in a metric space $(\calX, \mathrm{dist})$ with doubling dimension $\mathbf{ddim}$ and aspect ratio $\Delta$, there exists an algorithm that solves the $c$-ANN problem for $c = \frac{\alpha+1}{\alpha-1}+\eps$, using $\O{(4\alpha)^{\mathbf{ddim}} \log_\alpha \frac{\Delta}{(\alpha-1)\eps} \log \Delta}$ query time and $\O{n(4\alpha)^{\mathbf{ddim}} \log \Delta}$ space.
\end{theorem}

\section{Missing Proofs in \texorpdfstring{\secref{sec:color:sampling}}{Section 2}}
\seclab{sec:color:sampling:proof}

In this section, we provide the missing proofs for our random color sampling scheme in \secref{sec:color:sampling}.
We restate our \algref{alg:div:group:testing} in  \algref{alg:div:group:testing:back} for completeness.

\begin{algorithm}[!ht]
\caption{Restatement of \algref{alg:div:group:testing}}
\alglab{alg:div:group:testing:back}
\begin{algorithmic}[1]
\Require{Input dataset $P = \{p_1, \cdots, p_n\}$, each with a color $c \in \calC$, an integer $k \le |\calC|$}
\Ensure{Data structure to solve the $k$-diverse nearest neighbor problem}
\State{Let $L \gets \O{k \log k}$}
\For{$i \in [L]$}
\State{Sample each color $c \in \calC$ with probability $\frac{1}{k}$}
\State{Let $\calC_i$ be the set of sampled colors, and let $P_i$ be the set of points with colors in $\calC_i$}
\State{Maintain a data structure that solves the $c$-ANN problem for $P_i$} \Comment{Choose a subroutine from \thmref{thm:cnn:hamming}, \thmref{thm:cnn:euclidean}, \thmref{thm:cnn:metric:old}, \thmref{thm:cnn:metric} based on the dataset}
\EndFor
\For{each query $q$}
\State{Let $\widehat{p_i}$ be the reported $c$-ANN of bucket $i$ for each $i \in [L]$}
\State{{\bf Return:} $k$ nearest points to $q$ in $\{\widehat{p_1} \ldots, \widehat{p_L}\}$ of distinct colors}
\EndFor
\end{algorithmic}
\end{algorithm}

The following lemma shows the correctness of our randomized sampling procedure.

\begin{lemma}
\lemlab{lem:cor:group:testing}
\algref{alg:div:group:testing:back} solves the $k$-diverse $c$-ANN problem (see \defref{def:k:div:NN}) with probability $0.99$. 
\end{lemma}
\begin{proof}
Let $S \subset P$ be the set of $k$ nearest points to $q$ of distinct colors that satisfies $\mathrm{dist}(S,q) \le r$. 
Let $\{\widehat{p_1} \ldots, \widehat{p_L}\}$ be the set of reported nearest neighbors in each bucket of \algref{alg:div:group:testing:back}.
It suffices to show that there exists a set $S'$ of $k$ points of distinct colors in $\{\widehat{p_1} \ldots, \widehat{p_L}\}$ such that $\mathrm{dist}(S',q) \le cr$, as the correctness of \algref{alg:div:group:testing:back} follows. 
We prove this statement by an inductive argument.

For the initial stage, note that for each bucket, we sample each color $c \in \calC(S)$ with probability $\frac{1}{k}$. 
Here, we use $\calC(E)$ to denote the set of colors obtained by the points in a set $E \subset P$.
Since $|\calC(S)| = k$, a bucket contains a color $c \in \calC(S)$ with probability at least 
\[1-\left(1-\frac{1}{k}\right)^k \ge 1 - \frac{1}{e} = \Omega(1).\]
We suppose that bucket $i_1$ contains a color in $\calC(S)$, and let $T_1 = \{ \widehat{p_{i_1}} \}$, where $\widehat{p_{i_1}}$ is the reported nearest neighbor in bucket $i_1$.
Since bucket $i_1$ has color $c$, by our algorithm construction, it contains all points of color $c$, including the point in $S$.
Therefore, the nearest neighbor in bucket $i_1$ is within distance $r$ to $q$, and then if the $c$-ANN algorithm in bucket $i_1$ succeeds, which is with probability $0.99$, we have $\mathrm{dist}(T_1,q) \le cr$.
Then let event $\calE_1$ denote the event that a bucket contains a color $c \in \calC(S)$ and the $c$-ANN algorithm succeeds, we have $\PPr{\calE_1} = \Omega(1)$.
Let $X_1$ be the number of buckets needed to ensure that such a bucket $i_1$ exists, we have $\Ex{X_1} = \O{1}$.

Next, we suppose that there is a set $T_{j-1} \subset \{\widehat{p_1} \ldots, \widehat{p_L}\}$ such that it contains $j-1 < k$ points of distinct colors and satisfies $\mathrm{dist}(T_{j-1},q) \le cr$.
Then a bucket contains no color in $\calC(T_{j-1})$ with probability at least
\[\left(1-\frac{1}{k}\right)^{j-1} \ge \left(1-\frac{1}{k}\right)^k \ge \frac{1}{4} = \Omega(1).\]
In addition, since $|\calC(S \backslash T_{j-1})| = k-j+1$, a bucket contains a color in $\calC(S \backslash T_{j-1})$ with probability at least 
;\[ 1-\left(1-\frac{1}{k}\right)^{k-j+1} = 1 - e^{-\frac{k-j+1}{k}} \ge \frac{k-j+1}{2k} = \Omega(1) \cdot \frac{k-j+1}{k}. \]
We suppose that bucket $i_{j}$ contains a color in $\calC(S \backslash T_{j-1})$ and no colors in $T_{j-1}$, and let $T_{j} = T_{j-1} \cup \{ \widehat{p_{i_{j}}} \}$, where $\widehat{p_{i_{j}}}$ is the reported nearest neighbor in bucket $i_{j}$.
Again, since bucket $i_{j}$ selects the color $c$, by our algorithm construction, it contains all points of the color $c$, including the point in $S$.
Therefore, the nearest neighbor in bucket $i_{j}$ is within distance $r$ to $q$, and then if the $c$-ANN algorithm in bucket $i_{j}$ succeeds, which is with probability $0.99$, we have $\mathrm{dist}(T_{j},q) \le cr$.
Moreover, since bucket $i_{j}$ contains no colors in $\calC(T_j)$, $\widehat{p_{i_{j}}}$ has a new color, and so $T_{j}$ has $j$ distinct colors.
Then let event $\calE_j$ denote the event that a bucket contains a color $c \in \calC(S\backslash T_j)$ and no colors in $T_j$, and the $c$-ANN algorithm succeeds, we have $\PPr{\calE_j} = \Omega(\frac{k-j+1}{k})$. 
Let $X_j$ be the number of buckets needed to ensure that such a bucket $i_{j}$ exists, we have $\Ex{X_j} = \O{\frac{k}{k-j+1}}$.

Conditioned on the event that there exists bucket $i_1, \ldots, i_k$ that satisfy event $\calE_1, \ldots, \calE_k$ with $T_j := \{\widehat{p_{i_{j'}}}, j' \in [j]\}$ for each $j \in [k]$, then the above induction step shows that $T_k =  \{\widehat{p_{i_{j}}}, j \in [k]\} \subset \{\widehat{p_1} \ldots, \widehat{p_L}\}$ contains $k$ points of distinct colors and satisfies $\mathrm{dist}(T_k,q) \le cr$, demonstrating the desired statement.
Note that for each $j \in [k]$, we require $\O{\frac{k}{k-j+1}}$ buckets in expectation to ensure the existence of bucket $i_{j}$, which in total gives
\[\sum_{j \in [k]} \O{\frac{k}{k-j+1}} = \O{k \log k}\]
buckets in expectation.
Then, by Markov's inequality, with probability $0.99$, we need $\O{k \log k}$ buckets in total to guarantee the correctness.
Since we set $L = \O{k \log k}$ in \algref{alg:div:group:testing:back}, its output is correct with probability $0.99$.
\end{proof}

We next analyze the space and query time of our algorithm in Euclidean spaces.
\begin{theorem}[$k$-diverse $c$-ANN in Euclidean spaces]
\thmlab{thm:k:div:euclidean}
Given a set $P$ of $n$ points in the Euclidean space $(\mathbb{R}^d, \ell_2)$, there is an algorithm that solves the $k$-diverse $c$-ANN problem (see \defref{def:k:div:NN}) using $\O{k^{1-\frac{1}{c^2}-o(1)}\log k \cdot dn^{\frac{1}{c^2}+o(1)}}$ query time and $\O{\log k \cdot dn^{1+\frac{1}{c^2}+o(1)}}$ space with probability $0.99$.
\end{theorem}
\begin{proof}
It suffices to bound the query time and space of \algref{alg:div:group:testing:back} when we use the subroutine in \thmref{thm:cnn:euclidean} for $c$-ANN.
Now, we consider an arbitrary bucket.
For each color $c \in \calC$, let $n_c$ denote the number of points in $P$ of color $c$.
Let $X$ denote the number of points sampled in the bucket, we have $\Ex{X} = \frac{1}{k} \sum_{c \in \calC} n_c = \frac{n}{k}$.
By \thmref{thm:cnn:euclidean}, the $c$-ANN algorithm in each bucket uses $\O{dX^{\frac{1}{c^2}+o(1)}}$ query time and $\O{dX^{1+\frac{1}{c^2}+o(1)}}$ space.
Since $n^{\frac{1}{c^2}+o(1)}$ is a concave function, by Jensen's inequality, the expected query time in the bucket is at most 
\[\Ex{\O{dX^{\frac{1}{c^2}+o(1)}}} \le \O{d\Ex{X}^{\frac{1}{c^2}+o(1)}} \le \O{d\left(\frac{n}{k}\right)^{\frac{1}{c^2}+o(1)}}. \]
Then, since there are $L = \O{k \log k}$ buckets, the total query time in expectation is at most
\[L \cdot \O{d\left(\frac{n}{k}\right)^{\frac{1}{c^2}+o(1)}} = \O{k^{1-\frac{1}{c^2}-o(1)}\log k \cdot dn^{\frac{1}{c^2}+o(1)}}.\]
We note that there is an extra $\O{k \log k}$ factor in the query time due to the selection of the $k$ points of distinct colors from the reported points from all buckets, but it is dominated by the $\O{k^{1-\frac{1}{c^2}-o(1)}dn^{\frac{1}{c^2}+o(1)} \log k}$ factor.
In terms of space complexity, we write $X$ as $\sum_{c \in \calC} n_c X_c$, where $X_c$ is $1$ with probability $\frac{1}{k}$ and $0$ otherwise.
Note that $\Ex{d\left(\sum_{c \in \calC} n_c X_c\right)^{\frac{1}{c^2}+o(1)}}$ is Schur-convex in the vector $[n_1 \ldots, n_{|\calC|}]$, and it reaches the maximum when there is a $n_c = n$ while all others are $0$.
Then the expected space in each bucket is at most
\[\frac{1}{k}\cdot \O{dn^{1+\frac{1}{c^2}+o(1)}}.\]
Since there are $L = \O{k \log k}$ buckets, the total space in expectation is at most
\[L \cdot \frac{1}{k} \cdot \O{dn^{1+\frac{1}{c^2}+o(1)}} = \O{\log k \cdot dn^{1+\frac{1}{c^2}+o(1)}}. \]
By Markov's inequality, we have the desired result.
\end{proof}

An analogous guarantee holds for the Hamming metric, as stated below.

\begin{theorem}[$k$-diverse $c$-ANN with the Hamming metric]
\thmlab{thm:k:div:hamming}
Given a set $P$ of $n$ points in the Hamming metric space $(\{0,1\}^d, \ell_1)$, there is an algorithm that solves the $k$-diverse $c$-ANN problem (see \defref{def:k:div:NN}) using $\O{k^{1-\frac{1}{c}-o(1)}\log k \cdot dn^{\frac{1}{c}+o(1)}}$ query time and $\O{\log k \cdot dn^{1+\frac{1}{c}+o(1)}}$ space with probability $0.99$.
\end{theorem}
\begin{proof}
The analysis follows from the same calculation in \thmref{thm:k:div:euclidean}.
\end{proof}

In general metric spaces, combining our sampling approach with graph-based algorithms and navigating nets yields the following guarantees.

\begin{theorem}[$k$-diverse $c$-ANN with general metrics]
Given a set $P$ of $n$ points in a metric space $(\calX, \mathrm{dist})$ with doubling dimension $\mathbf{ddim}$ and aspect ratio $\Delta$, there is an algorithm that solves the $k$-diverse $c$-ANN problem (see \defref{def:k:div:NN}) for $c=\frac{\alpha+1}{\alpha-1}+\eps$ using 
$\O{k\log k \cdot (4\alpha)^{\mathbf{ddim}} \log_\alpha \frac{\Delta}{(\alpha-1)\eps} \log \Delta}$ query time and $\O{\log k\cdot n(4\alpha)^{\mathbf{ddim}} \log \Delta}$ space with probability $0.99$.
There is another algorithm that solves the $k$-diverse $c$-ANN problem using 
$\O{k\log k \cdot \left(2^{\O{\mathbf{ddim}}} \log \Delta + \left(\frac{1}{c-1}\right)^{\O{\mathbf{ddim}}}\right)}$ query time and $\O{\log k\cdot n2^{\O{\mathbf{ddim}}}}$ space with probability $0.99$.
\end{theorem}
\begin{proof}
We consider an arbitrary bucket.
For each color $c \in \calC$, let $n_c$ denote the number of points in $P$ of color $c$.
Let $X$ denote the number of points sampled in the bucket, we have $\Ex{X} = \frac{1}{k} \sum_{c \in \calC} n_c = \frac{n}{k}$.
By \thmref{thm:cnn:metric}, the $c$-ANN algorithm in each bucket uses $\O{(4\alpha)^{\mathbf{ddim}} \log_\alpha \frac{\Delta}{(\alpha-1)\eps} \log \Delta}$ query time and $\O{X(4\alpha)^{\mathbf{ddim}} \log \Delta}$ space.
Since there are $L = \O{k \log k}$ buckets, the total query time is $\O{k\log k \cdot (4\alpha)^{\mathbf{ddim}} \log_\alpha \frac{\Delta}{(\alpha-1)\eps} \log \Delta}$.
The expected space in each bucket is $\O{\Ex{X}(4\alpha)^{\mathbf{ddim}} \log \Delta} = \O{\frac{n}{k}(4\alpha)^{\mathbf{ddim}} \log \Delta}$, and so we in total need $\O{\log k \cdot n(4\alpha)^{\mathbf{ddim}} \log \Delta}$ space in expectation.
By Markov's inequality, we have the desired result.
The second result follows from a similar proof, where we instead apply the $c$-ANN algorithm in \thmref{thm:cnn:metric:old} to each bucket.
\end{proof}

\paragraph{A relaxed definition.} 
We consider a relax definition of the $k$-diverse problem called the $(k,\eps)$-diverse problem, where we allow to have only $(1-\eps)\cdot k$ points of distinct colors in the solution set of $k$ points.
This indicates that the recommendation system could return redundant or replicated results within distance $cr$ to the query point, as long as there are $(1-\eps)\cdot k$ diverse points.
The formal definition is stated as follows:

\begin{definition}[The $(k,\eps)$-diverse $c$-ANN problem]
\deflab{def:k:eps:div:NN}
Given a set $P$ of $n$ points in a metric space $(\calX, \mathrm{dist})$ and parameter $r > 0$, the goal is to construct a data structure that satisfies the following: given any query $q$, if the smallest ball around $q$ which contains $k$ points of distinct colors has radius at most $r$, it returns a set $\widehat{S}$ of $(1-\eps)\cdot k$ points of distinct colors that satisfies $\max_{p \in \widehat{S}}\dist(p,q) \le cr$. 
\end{definition}

We show that our color sampling approach still applies to the relaxed setting, and moreover, we replace the $\log k$ dependence by $\log \frac{1}{\eps}$, achieving more efficient query time.

\begin{lemma}
\lemlab{lem:cor:group:testing:relaxed}
Implementing \algref{alg:div:group:testing:back} with $L = \O{k \log \frac{1}{\eps}}$ solves the $(k,\eps)$-diverse $c$-ANN problem (see \defref{def:k:eps:div:NN}) with probability $0.99$. 
\end{lemma}
\begin{proof}
Without loss of generality, we assume $\eps k$ is an integer. 
Let $S \subset P$ be the set of $k$ nearest points to $q$ of distinct colors that satisfies $\mathrm{dist}(S,q) \le r$. 
Let $\{\widehat{p_1} \ldots, \widehat{p_L}\}$ be the set of reported nearest neighbors in each bucket of \algref{alg:div:group:testing:back}.
It suffices to show that there exists a set $S'$ of $(1-\eps)\cdot k$ points of distinct colors in $\{\widehat{p_1} \ldots, \widehat{p_L}\}$ such that $\mathrm{dist}(S',q) \le cr$.

Note that in the proof of \lemref{lem:cor:group:testing}, we show that for any set $T_{j-1} \subset \{\widehat{p_1} \ldots, \widehat{p_L}\}$ such that it contains $j-1 < k$ points of distinct colors and satisfies $\mathrm{dist}(T_{j-1},q) \le cr$, \algref{alg:div:group:testing:back} reports a nearest neighbor with a new color within radius $cr$ using additional $\O{\frac{k}{k-j+1}}$ buckets in expectation.
Therefore, to reach $(1-\eps)\cdot k$ colors we only need
\[\sum_{j \in [(1-\eps)\cdot k]} \O{\frac{k}{k-j+1}} = \O{k \log \frac{1}{\eps}}\]
buckets in expectation.
Thus, by Markov's inequality, using $L = \O{k \log \frac{1}{\eps}}$ buckets in \algref{alg:div:group:testing:back} gives us the desired result with probability $0.99$.
\end{proof}

The following theorem states the guarantees of our $(k,\eps)$-diverse $c$-ANN algorithm in Euclidean spaces.
The analysis follows directly from that of the $k$-diverse $c$-ANN algorithms (see e.g., \thmref{thm:k:div:euclidean}) and hence is omitted.

\begin{theorem}[$(k,\eps)$-diverse $c$-ANN in Euclidean spaces]
Given a set $P$ of $n$ points in the Euclidean space $(\mathbb{R}^d, \ell_2)$, there is an algorithm that solves the $(k,\eps)$-diverse $c$-ANN problem (see \defref{def:k:eps:div:NN}) using $\O{k^{1-\frac{1}{c^2}-o(1)}\log \frac{1}{\eps} \cdot dn^{\frac{1}{c^2}+o(1)}}$ query time and $\O{\log \frac{1}{\eps} \cdot dn^{1+\frac{1}{c^2}+o(1)}}$ space with probability $0.99$.
\end{theorem}

The following theorem states the guarantees of the Hamming metric.

\begin{theorem}[$(k,\eps)$-diverse $c$-ANN with the Hamming metric]
Given a set $P$ of $n$ points in the Hamming metric space $(\{0,1\}^d, \ell_1)$, there is an algorithm that solves the $(k,\eps)$-diverse $c$-ANN problem (see \defref{def:k:eps:div:NN}) using $\O{k^{1-\frac{1}{c}-o(1)}\log \frac{1}{\eps} \cdot dn^{\frac{1}{c}+o(1)}}$ query time and $\O{\log \frac{1}{\eps} \cdot dn^{1+\frac{1}{c}+o(1)}}$ space with probability $0.99$.
\end{theorem}

The following theorem states the guarantees of general metrics.

\begin{theorem}[$(k,\eps)$-diverse $c$-ANN with general metrics]
Given a set $P$ of $n$ points in a metric space $(\calX, \mathrm{dist})$ with doubling dimension $\mathbf{ddim}$ and aspect ratio $\Delta$, there is an algorithm that solves the $(k,\eps)$-diverse $c$-ANN problem (see \defref{def:k:eps:div:NN}) using 
$\O{k\log \frac{1}{\eps} \cdot (4\alpha)^{\mathbf{ddim}} \log_\alpha \frac{\Delta}{(\alpha-1)\eps} \log \Delta}$ query time and $\O{\log \frac{1}{\eps}\cdot n(4\alpha)^{\mathbf{ddim}} \log \Delta}$ space with probability $0.99$.
There is another algorithm that solves the $(k,\eps)$-diverse $c$-ANN problem using 
$\O{k\log \frac{1}{\eps} \cdot \left(2^{\O{\mathbf{ddim}}} \log \Delta + \left(\frac{1}{c-1}\right)^{\O{\mathbf{ddim}}}\right)}$ query time and $\O{\log \frac{1}{\eps}\cdot n2^{\O{\mathbf{ddim}}}}$ space with probability $0.99$.
\end{theorem}

\section{Algorithms with Redundant Points}
\seclab{sec:red}

In this section, we consider another relaxed notion of the diverse problem studied by \cite{AnandIKMRSX25} called the $(k,k')$-diverse problem.
In this setting, we need to report $k$ nearest neighbors in total and allow each color in the reported points to have $k'$ redundant points.
We now formalize the definition of $(k,k')$-diverse.

\begin{definition}[$(k,k')$-diverse]
\deflab{def:k':nn}
Given a set $P$ of $n$ points in a metric space $(\calX, \mathrm{dist})$, where each point is assigned a color. Let $S$ be a set of points in $X$, we say a subset $S\subset P$ is $(k,k')$-diverse if $|S|=k$ and for any point $p \in S$, there are at most $k'$ points in $S$ that have the same color as $P$. 
Note that being $(k, 1)$-diverse is equivalent to the
standard notion of $k$-diverse.
\end{definition}

The following definition $(k,k')$-diverse $c$-ANN problem is then induced by the above notion.

\begin{definition}[The $(k,k')$-diverse $c$-ANN problem]
\deflab{def:k:k':div:NN}
Given a set $P$ of $n$ points in a metric space $(\calX, \mathrm{dist})$, where each point is assigned a color, and parameter $r > 0$, the goal is to construct a data structure that satisfies the following: given any query $q$, if the smallest ball around $q$ which contains a $(k,k')$-diverse subset of $P$ has radius at most $r$, it returns a set $\widehat{S}$ that satisfies $\max_{p \in \widehat{S}}\dist(p,q) \le cr$ and is $(k,k')$-diverse.
\end{definition}

Recall that we sample each color with probability $\frac{1}{k}$ in our color sampling scheme in \algref{alg:div:group:testing}.
As we allow to report $k'$ points for each color, we instead sample each color proportional to $\frac{k'}{k}$ and report $k'$ nearest neighbors from each bucket.
Similarly, we can show that the candidate sets from all buckets collectively recover a set of $k$ nearest neighbors satisfying the $(k,k')$-diverse requirement, giving us a valid approximate solution to the $(k,k')$-diverse problem.
We formally summarize this process in \algref{alg:div:group:testing:redundant}.
To begin with, we introduce a modified version of the LSH that returns $m$ nearest neighbors of a query, so that we can apply in each bucket.
\begin{theorem}
\thmlab{thm:cnn:euclidean:k}
Given a set $P$ of $n$ points in the Euclidean space $(\mathbb{R}^d, \ell_2)$, a query $q$, and an integer $m$, there exists an algorithm that returns a set $\widetilde{S}$ of $m$ points.
Let $r$ be the distance between $q$ and its nearest neighbor in $P \backslash \widetilde{S}$, then for each point $p \in \widetilde{S}$, $\dist(p,q) \le cr$ with probability $0.99$.
The algorithm uses $\O{mdn^{\frac{1}{c^2}+o(1)}}$ query time and $\O{dn^{1+\frac{1}{c^2}+o(1)}}$ space and preprocessing.
\end{theorem}
\begin{proof}
Recall that in the $c$-ANN algorithm in \thmref{thm:cnn:euclidean:k}, we hash each point to several buckets using LSH.
Given an LSH object, we report $m$ nearest neighbors iteratively: Whenever we return a point by the $c$-ANN algorithm, we delete the returned point from all the LSH buckets it hashes into and continue to find the next with the new LSH object.
Since each point is hashed independently in LSH, in the $t$-the round of the iterative process, the LSH data structure that we maintain is equivalent to the LSH implemented in the dataset $P \backslash \{\widetilde{p_1}, \ldots, \widetilde{p_t}$, where $p_\tau$ is the point we obtained in round $\tau \in [t]$.
Let $p^*$ be the nearest neighbor of $q$ in $P \backslash \widetilde{S}$, then $p^*$ always exists in the LSH data structure $m$ rounds.
Conditioned on the success of each query of the LSH object, the point we returned in each round is always a $c$-ANN of $p^*$ that has distance $r$ to $q$, giving the guarantee in \thmref{thm:cnn:euclidean:k}.
Finally, notice that we query $m$ times the LSH data structure with at most $n$ points and the time to delete the point in each round is dominated by the query time, the total time we need is then $m$ times that of \thmref{thm:cnn:euclidean:k}.
Note that the iterative procedure does not incur additional space, so the space follows directly from \thmref{thm:cnn:euclidean:k}.
\end{proof}


\begin{algorithm}[!ht]
\caption{Diversity search allowing redundant points}
\alglab{alg:div:group:testing:redundant}
\begin{algorithmic}[1]
\Require{Input dataset $P = \{p_1, \cdots, p_n\}$, each with a color $c \in \calC$, an integer $k \le |\calC|$}
\Ensure{Data structure to solve the $(k,k')$-diverse nearest neighbor problem}
\State{Let $L \gets \O{\frac{k \log k}{k'}}$}
\For{$i \in [L]$}
\State{Sample each color $c \in \calC$ with probability $\frac{k'}{2k}$}
\State{Let $\calC_i$ be the set of sampled colors, and let $P_i$ be the set of points with colors in $\calC_i$}
\State{Maintain a data structure that reports $k'$ $c$-ANNs of each query for $P_i$} \Comment{Choose a subroutine from \thmref{thm:cnn:hamming}, \thmref{thm:cnn:euclidean}, \thmref{thm:cnn:metric:old}, \thmref{thm:cnn:metric} based on the dataset}
\EndFor
\For{each query $q$}
\State{Let $\widetilde{S_i}$ be the set of the reported $k$ $c$-ANNs of bucket $i$ for each $i \in [L]$}
\State{$\calL \gets \emptyset$}
\For{each point $h$ in $\bigcup_i \widetilde{p_i}$ sorted by the distance to $q$}
\State{$\calL \gets \calL \cup \{h\}$ if $\calL \cup \{h\}$ is $(|\calL|+1,k')$-diverse}\Comment{There are less than $k'$ points in $\calL$ that have the same color as $h$}
\EndFor
\State{{\bf Return:} the list $\calL$}
\EndFor
\end{algorithmic}
\end{algorithm}

We next show the correctness of \algref{alg:div:group:testing:redundant}.
\begin{lemma}
\lemlab{lem:div:group:testing:redundant}
\algref{alg:div:group:testing:redundant} solves the $(k,k')$-diverse $c$-ANN problem (see \defref{def:k:k':div:NN}) with probability $0.99$. 
\end{lemma}
\begin{proof}
Let $S \subset P$ be a $(k,k')$-diverse subset that satisfies $\mathrm{dist}(S,q) \le r$. 
Let $\widetilde{S_1} \ldots, \widetilde{S_L}$ be the set of reported nearest neighbors in each bucket of \algref{alg:div:group:testing:redundant}.
It suffices to show that there exists a $(k,k')$-diverse subset in $\bigcup_i \widetilde{p_i}$ such that $\mathrm{dist}(S',q) \le cr$, as the correctness of \algref{alg:div:group:testing:redundant} follows from the greedy search in the post-processing part. 
We prove this statement by a similar inductive argument as in \lemref{lem:cor:group:testing}.
We construct a candidate solution set $S'$ gradually by adding $\frac{k'}{4}$ points at each step while preserving the following invariants: (1). $\dist(S,q) \le cr$ and (2). $S'$ is $(|S|,k')$-diverse.
We maintain a list of colors $\calC_{\mathrm{bad}}$ that contains all colors $c \notin \calC(S)$ such that there are $\ge \frac{k'}{4}$ points in $S'$ with color $c$.
Here, we use $\calC(E)$ to denote the set of colors obtained by the points in a set $E \subset P$.
We also maintain a list of points $S_\mathrm{valid}$ that contains all points with colors in $\calS(S') \backslash \left( \calC(S) \cup \calC_{\mathrm{bad}}\right)$, i.e., they are selected to the candidate solution set $S'$ but they are neither in the optimal solution set $S$ nor a ``bad'' color.
We consider $k = \beta \cdot \frac{k'}{4} + \gamma$ for some integer $\beta$ and $0 \le \gamma \le \frac{k'}{4}$.

For the initial stage, note that we sample each color $c \in \calC(S)$ with probability $\frac{k'}{2k}$ independently, so the expected number of points sampled in the optimal solution $S$ is $\frac{k'}{2}$. 
Therefore, a bucket contains at least $\frac{k'}{4}$ points in $S$ with probability $\Omega(1)$ by the property of Bernoulli variables.
We suppose that bucket $i_1$ is such a bucket and consider the following cases.
In the first case, the $2k'$ $c$-ANNs of bucket $i_1$, denoted by $\widetilde{S_1}$, contain all the $\frac{k'}{4}$ points in $S$, then we add these points to $S'$.
If this case does not hold, then at least one point in the optimal solution $S$ is not included in $\widetilde{S_1}$, and by \thmref{thm:cnn:euclidean:k}, all the $2k'$ reported points have distance $\le cr$ to the query $q$ conditioned on the success of the $c$-ANN algorithms.
Next, in the second case, there is a color that has at least $\frac{k'}{2}$ points in the reported points, then we add the closest $\frac{k'}{4}$ points to $S$.
Note that the $c$-ANN algorithm succeeds with probability $0.99$, and so with probability $\Omega(1)$, the closest $\frac{k'}{4}$ points in the $\frac{k'}{2}$ points mentioned in the previous sentence are correct $c$-ANNs.
If neither case one nor case two occurs, we add the closest $\frac{k'}{4}$ points in $\widetilde{S_1}$ to $S$.
In all cases, the invariants are automatically satisfied.
Then let event $\calE_1$ denote the event that the invariants hold after the first step, we have $\PPr{\calE_1} = \Omega(1)$.
So, let $X_1$ be the number of independent buckets needed to ensure that such a bucket $i_1$ exists, we have $\Ex{X_1} = \O{1}$.

Next, we suppose that after the $(j-1)$-th step, the invariants hold.
Consider the $j$-th step, at the beginning, there is $\frac{(j-1)k'}{4}$ points in $S'$, and so there are at least $k - \frac{(j-1)k'}{4}$ points in $S$ that have not been added to $S'$.
For each bucket, the expected number of points sampled from $S \backslash S'$ is $\frac{k'}{2k} \cdot \left(k - \frac{(j-1)k'}{4}\right)$, and by the property of Bernoulli variables, we sample $\frac{k'}{4}$ points with $S \backslash S'$ with probability $\Omega\left(\frac{k-jk'}{k}\right)$. 
Moreover, consider the ``bad'' color list $\calC_\mathrm{bad}$, its size is less than $j-1$ since each color inside has at least $\frac{k'}{4}$ points in $S'$ and there are $\frac{(j-1)k'}{4}$ points in $S'$ at the current step.
Then a bucket contains no color in $\calC_\mathrm{bad}$ with probability at least
\[\left(1-\frac{k'}{2k}\right)^{j-1} \ge \left(1-\frac{k'}{2k}\right)^{k} \ge \frac{1}{16} = \Omega(1).\]
In addition, consider the ``valid'' point list $S_\mathrm{valid}$, we have $|S_\mathrm{valid}| \le k$ because $S_\mathrm{valid} \subset S$.
Then the expected number of points in $S_\mathrm{valid}$ sampled to each bucket is $\frac{k'}{2}$, and so by Markov's inequality, each bucket contains at most $\frac{3k'}{4}$ points in $S_\mathrm{valid}$ with probability $\Omega(1)$.
Now notice that $\cal(S \backslash S')$, $\calC_\mathrm{bad}$, $\calC(S_\mathrm{valid})$ are disjoint sets, so the three events we discussed above are independent, since we sample each color independently.
Thus, let event $\calE_j$ denote the event that a bucket contains at least $\frac{k'}{4}$ points from $ S \backslash S'$, no colors from $\calC_\mathrm{bad}$, and at most $\frac{3k'}{4}$ points from $S_\mathrm{valid}$, event $\calE_j$ is satisfied with probability $\Omega\left(\frac{k-jk'}{k}\right)$.
Let $X_j$ be the number of buckets needed to ensure that such a bucket $i_{j}$ exists, we have $\Ex{X_j} = \O{\frac{k}{k-jk'}}$.

We suppose that bucket $i_{j}$ is a bucket that ensures event $\calE_j$.
Again, in the first case, the returned set of $2k'$ $c$-ANNs $\widetilde{S_{j}}$ by bucket $i_j$ contains at least $\frac{k'}{4}$ points in $S \backslash S'$, then we add these points to $S'$.
Note that the invariants are satisfied since these points are included in the optimal solution set $S$ and have not been added to $S'$ before. 
If the first case does not happen, conditioned on the correctness of the $c$-ANN algorithm, all points in $\widetilde{S_{j-1}}$ have distance $\le cr$ to the query $q$.
Next, in the second case, there is a color $c$ in $\calC(\widetilde{p_{j}})$ that has at least $\frac{3k'}{4}$ points.
Since we do not have colors in $\calC_\mathrm{bad}$ in bucket $i_j$, there are at most $\frac{k'}{4}$ points in $S'$ with color $c$, and thus, there are at least $\frac{k'}{2}$ points in $\widetilde{S_{j-1}} \backslash S'$ with color $c$.
Therefore, we add the closest $\frac{k'}{4}$ points to $S'$ without violating the invariants.
Similarly, these $\frac{k'}{4}$ points are valid $c$-ANNs with probability $\Omega(1)$.
Finally, if none of the above cases occur, then since $\calC(\widetilde{p_{j}})$ has at most $\frac{k'}{4}$ points in $S \backslash S'$ and at most $\frac{3k'}{4}$ points in $S'$, there are at least $k'$ points in $\widetilde{S_{j}}$ that are not in both $S$ and $S'$.
Moreover, we can add the closest $\frac{k'}{4}$ points of these $k'$ points to $S'$ without violating the $(k,k')$-diverse requirement, because none of the attained colors in $\calC(\widetilde{p_{j}})$ has at least $\frac{k'}{2}$ points in $\widetilde{S_{j}}$ and at least $\frac{k'}{4}$ in $S'$ as there are no ``bad'' colors.
Lastly, since the closest $\frac{k'}{4}$ points mentioned in the previous sentence are within distance $cr$ to $q$ with probability $\Omega(1)$, we do not violate the ANN requirement, and so the invariants hold.

Note that the above argument only holds for $k_\mathrm{rem} = \Omega(k')$, where $k_\mathrm{rem} := k - \frac{(j-1)k'}{4}$ is defined as the number of points we need in the candidate solution set, as we cannot obtain $\frac{k'}{4}$ in one bucket when $k$ is smaller than $\Theta(k')$.
However, for each bucket, the expected number of sampled points is $\frac{k'}{2k} \cdot k_\mathrm{rem}$, and so we sample at least $\frac{k_\mathrm{rem}}{2}$ points with probability $\Omega\left(\frac{k}{k'}\right)$.
Then with the same analysis, let event $\calE_j$ denote the event that a bucket contains at least $\frac{k_\mathrm{rem}}{2}$ points from $ S \backslash S'$, no colors from $\calC_\mathrm{bad}$, and at most $\frac{3k'}{4}$ points from $S_\mathrm{valid}$, event $\calE_j$ is satisfied with probability $\Omega\left(\frac{k'}{k}\right)$.
Let $X_j$ be the number of buckets needed to ensure that such a bucket $i_{j}$ exists, we have $\Ex{X_j} = \O{\frac{k}{k'}}$.
Now we consider bucket $i_j$ that ensures event $\calE_j$.
Under the first case, we can add the $\frac{k_\mathrm{rem}}{2}$ points in $S \backslash S'$ to $S'$; and under the second case and the third case, we can add $\frac{k'}{4} = \Omega\left(\frac{k_\mathrm{rem}}{2}\right)$ points to $S'$.
Thus, $k_\mathrm{rem}$ decreases by at least half at each step, and so there are at most $\O{\log k'}$ steps before we obtain a candidate set with size $k$, where we end the iteration.

Since in each step the two invariants hold, the candidate set suffices to give a valid $(k,k')$-diverse $c$-ANN solution.
We now compute the number of buckets needed.
For $k_\mathrm{rem} > \Omega(k')$, we need $\Ex{X_j} = \O{\frac{k}{k-jk'}}$ buckets in step $j$ to recover a valid bucket; and for $k_\mathrm{rem} \le \Omega(k')$, we need $\Ex{X_j} = \O{\frac{k}{k'}}$ buckets each step and there are at most $\O{\log k'}$ steps.
Therefore, we in total need
\[\sum_{j \in [\beta]} \O{\frac{k}{k-jk'}} + \O{\log k'}\cdot \O{\frac{k}{k'}} = \O{\frac{k \log k}{k'}}\]
buckets in expectation, where $\beta = \O{\frac{k}{k'}}$.
By Markov's inequality, with probability $0.99$, we need $\O{\frac{k \log k}{k'}}$ buckets to guarantee the correctness.
Since we set $L = \O{\frac{k \log k}{k'}}$ in \algref{alg:div:group:testing:redundant}, its output is correct with probability $0.99$.
\end{proof}

Combining the framework in \algref{alg:div:group:testing:redundant} with the LSH-based $c$-ANN algorithm gives us the following result:
\begin{theorem}[$k$-diverse $c$-ANN in Euclidean spaces]
\thmlab{thm:k:div:euclidean:k'}
Given a set $P$ of $n$ points in the Euclidean space $(\mathbb{R}^d, \ell_2)$, there is an algorithm that solves the $k$-diverse $c$-ANN problem (see \defref{def:k:div:NN}) using $\O{k^{1-\frac{1}{c^2}-o(1)} (k')^{\frac{1}{c^2}+o(1)}\log k \cdot dn^{\frac{1}{c^2}+o(1)}}$ query time and $\O{\log k \cdot dn^{1+\frac{1}{c^2}+o(1)}}$ space with probability $0.99$.
\end{theorem}
\begin{proof}
The analysis is similar to \thmref{thm:k:div:euclidean:k'}.
It suffices to bound the query time and space of \algref{alg:div:group:testing:redundant} when we use the subroutine in \thmref{thm:cnn:euclidean:k} for $c$-ANN.
Now, we consider an arbitrary bucket.
For each color $c \in \calC$, let $n_c$ denote the number of points in $P$ of color $c$.
Let $X$ denote the number of points sampled in the bucket, we have $\Ex{X} = \frac{k'}{k} \sum_{c \in \calC} n_c = \frac{nk'}{k}$.
By \thmref{thm:cnn:euclidean:k}, the $c$-ANN algorithm in each bucket uses $\O{k'X^{\frac{1}{c^2}+o(1)}}$ query time and $\O{dX^{1+\frac{1}{c^2}+o(1)}}$ space since we return $k'$ points in each bucket.
Since $n^{\frac{1}{c^2}+o(1)}$ is a concave function, by Jensen's inequality, the expected query time in the bucket is at most 
\[\Ex{\O{dk'X^{\frac{1}{c^2}+o(1)}}} \le \O{dk'\Ex{X}^{\frac{1}{c^2}+o(1)}} \le \O{dk'\left(\frac{nk'}{k}\right)^{\frac{1}{c^2}+o(1)}}. \]
Then, since there are $L = \O{\frac{k \log k}{k'}}$ buckets, the total query time in expectation is at most
\[L \cdot \O{dk'\left(\frac{nk'}{k}\right)^{\frac{1}{c^2}+o(1)}} = \O{k^{1-\frac{1}{c^2}-o(1)}(k')^{\frac{1}{c^2}+o(1)}\log k \cdot dn^{\frac{1}{c^2}+o(1)}}.\]
We note that there is an extra $\O{k \log k}$ factor in the query time due to the selection of the $(k,k')$-diverse subset from the reported points from all buckets, but it is dominated by the other factor.
Moreover, by \thmref{thm:cnn:euclidean:k} we know that the space to implement the $c$-ANN algorithm in each bucket does not incur additional space, so the space complexity direct follows from \thmref{thm:k:div:euclidean}.
\end{proof}

\section{Missing Proofs in \texorpdfstring{\secref{sec:expander}}{Section 3}}
\seclab{sec:expander:proof}

In this section, we provide the missing proofs for our expander graph-based color bucketing scheme in \secref{sec:expander}.
We first formalize our algorithm in \algref{alg:div:expander:back}.

\begin{algorithm}[!ht]
\caption{Formal statement of the algorithm in \secref{sec:expander}}
\alglab{alg:div:expander:back}
\begin{algorithmic}[1]
\Require{Input dataset $P = \{p_1, \cdots, p_n\}$, each with a color $c \in \calC$, an integer $k \le |\calC|:=N$, error ratio $\eps >0$}
\Ensure{Data structure to solve the $k$-diverse nearest neighbor problem}
\State{Let $\gamma = \frac{1}{\sqrt{\log k}}$, $D = \O{\left(\frac{\log k \log N}{\eps}\right)^{1+\frac{1}{\gamma}}}$ and $M = D^2 k^{1+\gamma}$}
\State{Maintain a $(k,(1-\eps)D)$-expander $G=(\calC,B,E)$ with $E: [|\calC|] \times [D] \to [M]$}
\Comment{$B =\{b_1,\ldots, b_{|M|}\}$ denotes the set of buckets}
\For{each vertex $b_i \in [B]$}
\State{Let $\calC_i$ contains the colors of all the right-neighbors of $b_i$, and let $P_i$ be the set of points with colors in $\calC_i$}
\State{Maintain a data structure that solves the $c$-ANN problem for $P_i$} \Comment{Choose a subroutine from \thmref{thm:cnn:metric:old} or \thmref{thm:cnn:metric} based on the dataset}
\EndFor
\For{each query $q$}
\State{Let $\widehat{p_i}$ be the reported $c$-ANN of bucket $i$ for each $i \in [L]$}
\State{{\bf Return:} $k$ nearest points to $q$ in $\{\widehat{p_1} \ldots, \widehat{p_L}\}$ of distinct colors}
\EndFor
\end{algorithmic}
\end{algorithm}

The next statement shows the correctness of \algref{alg:div:expander:back}.
\begin{lemma}
\lemlab{lem:cor:expander}
\algref{alg:div:expander:back} solves the $(k,\eps)$-diverse $c$-ANN problem (see \defref{def:k:eps:div:NN}). 
\end{lemma}
\begin{proof}
The analysis follows the framework in \lemref{lem:cor:group:testing}.
Let $S \subset P$ be the set of $k$ nearest points to $q$ of distinct colors that satisfies $\mathrm{dist}(S,q) \le r$. 
Let $\{\widehat{p_1} \ldots, \widehat{p_L}\}$ be the set of reported nearest neighbors in each bucket of \algref{alg:div:expander:back}.
It suffices to show that there exists a set $S'$ of $(1-\eps)\cdot k$ points of distinct colors in $\{\widehat{p_1} \ldots, \widehat{p_L}\}$ such that $\mathrm{dist}(S',q) \le cr$. 
Here, without loss of generality, we assume $(1-\eps)\cdot k$ is an integer.
We then prove this statement by an inductive argument.

We use $\calC(E)$ to denote the set of colors obtained by the points in a set $E \subset P$.
Let $G=(\calC,B,E)$ be the $(k,(1-\eps)D)$-expander obtained by \algref{alg:div:expander:back}.
For the initial stage, let bucket $i_1$ be a bucket that contains a color $c\in \calC(S)$, and let $T_1 = \{ \widehat{p_{i_1}} \}$, where $\widehat{p_{i_1}}$ is the reported nearest neighbor in bucket $i_1$.
Since bucket $i_1$ has color $c$, by our algorithm construction, it contains all points of color $c$, including the point in $S$.
Therefore, the nearest neighbor in bucket $i_1$ is within distance $r$ to $q$, and then the $c$-ANN algorithm in bucket $i_1$ gives $\mathrm{dist}(T_1,q) \le cr$.

Next, we suppose that there is a set $T_{j-1} \subset \{\widehat{p_1} \ldots, \widehat{p_L}\}$ such that it contains $j-1 < (1-\eps)\cdot k$ points of distinct colors and satisfies $\mathrm{dist}(T_{j-1},q) \le cr$.
Let $N(\calC(T_{j-1})\cup S)$ be the neighborhood of $\calC(T_{j-1})\cup S$ in $B$, it satisfies $|N(\calC(T_{j-1})\cup S)|\ge(1-\eps)\cdot kD$ by \thmref{thm:expander}.
Then since each color in $\calC(T_{j-1})$ only has $D$ neighbors and $|\calC(T_{j-1})| < (1-\eps)\cdot k$, there is at least one bucket in $N(\calC(T_{j-1})\cup S)$ that contains no colors in $\calC(T_{j-1})$.
In addition, since it is in the neighborhood of $\calC(T_{j-1})\cup S$, it must contain at least one color $c \in S \backslash \calC(T_{j-1})$.
Let $i_{j}$ denote this bucket and let $T_{j} = T_{j-1} \cup \{ \widehat{p_{i_{j}}} \}$, where $\widehat{p_{i_{j}}}$ is the reported nearest neighbor in bucket $i_{j}$.
Again, since bucket $i_{j}$ selects the color $c$, by our algorithm construction, it contains all points of the color $c$, including the point in $S$.
Therefore, the nearest neighbor in bucket $i_{j}$ is within distance $r$ to $q$, and then the $c$-ANN algorithm in bucket $i_{j}$ gives $\mathrm{dist}(T_{j},q) \le cr$.
Moreover, since bucket $i_{j}$ contains no colors in $\calC(T_j)$, $\widehat{p_{i_{j}}}$ has a new color, and so $T_{j}$ has $j$ distinct colors.
We iterate this inductive argument until $j=(1-\eps) \cdot k$, where we find at least $(1-\eps) \cdot k$ distinct colors in the reported set.
Thus, we prove the desired statement.
\end{proof}

We note that the diskANN algorithm studied by \cite{JSDS+19,IndykX23} is deterministic.
Thus, combining with our expander-based color bucketing approach, we obtain deteministic algorithms for the $(k,\eps)$-diverse $c$-ANN problem.

\begin{theorem}[$(k,\eps)$-diverse $c$-ANN on general metric spaces]
\thmlab{thm:k:div:det}
Given a set $P$ of $n$ points in a metric space $(\calX, \mathrm{dist})$ with doubling dimension $\mathbf{ddim}$, aspect ratio $\Delta$, and $N$ colors, there is an algorithm that solves the $(k,\eps)$-diverse $c$-ANN problem using 
\[\O{k^{1+o(1)}\left(\frac{\log k \log N}{\eps}\right)^{\O{\sqrt{\log k}}}\cdot(4\alpha)^{\mathbf{ddim}} \log_\alpha \frac{\Delta}{(\alpha-1)\eps} \log \Delta}\]
query time and 
\[\O{\left(\frac{\log k \log N}{\eps}\right)^{\O{\sqrt{\log k}}} \cdot n(4\alpha)^{\mathbf{ddim}} \log \Delta}.\]
space.
\end{theorem}
\begin{proof}
It suffices to compute the query time and space of \algref{alg:div:expander:back}.
We consider an arbitrary bucket $b$.
For each color $c \in \calC$, let $n_c$ denote the number of points in $P$ of color $c$.
Let $X$ denote the number of points in bucket $b$, we have $X = \sum_{c \in b} n_c$.
By \thmref{thm:cnn:metric}, the $c$-ANN algorithm in each bucket uses $\O{(4\alpha)^{\mathbf{ddim}} \log_\alpha \frac{\Delta}{(\alpha-1)\eps} \log \Delta}$ query time and $\O{X_b(4\alpha)^{\mathbf{ddim}} \log \Delta}$ space.
Recall that our expander $G=(\calC,B,E)$ has degree $D = \O{\left(\frac{\log k \log N}{\eps}\right)^{1+\frac{1}{\gamma}}}$ and number of buckets $M \le D^2 k^{1+\gamma}$ with $\gamma = \frac{1}{\sqrt{\log k}}$.
Then the total query time is $\O{M(4\alpha)^{\mathbf{ddim}} \log_\alpha \frac{\Delta}{(\alpha-1)\eps} \log \Delta}$, or equivalently
\begin{align*}
& \O{k^{1+o(1)}\left(\frac{\log k \log N}{\eps}\right)^{\O{\sqrt{\log k}}}\cdot(4\alpha)^{\mathbf{ddim}} \log_\alpha \frac{\Delta}{(\alpha-1)\eps} \log \Delta}.
\end{align*}
Moreover, the total space is $\O{\sum_{b \in B} X_b(4\alpha)^{\mathbf{ddim}} \log \Delta}$.
Then since each color $c$ has degree at most $D$, we have
\[\sum_{b \in B} X_b = \sum_{b \in B} \sum_{c \in b} n_c = \sum_{c \in \calC} D n_c = D\cdot n.\]
Thus, the total space is
\[\O{Dn(4\alpha)^{\mathbf{ddim}} \log \Delta} = \O{\left(\frac{\log k \log N}{\eps}\right)^{\O{\sqrt{\log k}}} \cdot n(4\alpha)^{\mathbf{ddim}} \log \Delta}.\]
\end{proof}

\section{Lower Bound for Dataset-Oblivious Bucketing}
\seclab{sec:lb}
In this section, we provide the formal discussion and analysis for our lower bound in \secref{sec:lb:front}.
Recall that in \secref{sec:expander}, we give a deterministic color bucketing method based on expander graphs, which provides a $(k,\eps)$-diverse solution.
Then a natural question is whether dataset-oblivious algorithms can give exact solutions.
In this section, we provide a lower bound on deterministic dataset-oblivious color bucketing schemes.
A bucketing scheme is dataset-oblivious if the subsets of colors contained in each bucket are fixed independently of the query and the spatial coordinates of the dataset. 
We demonstrate that such non-adaptive schemes require $\Omega\left(k^2\right)$ buckets.

\begin{definition}[Dataset-Oblivious Bucketing Scheme]
A deterministic color bucketing scheme is dataset-oblivious if the family of color buckets $\mathcal{B}=\left\{B_1, \ldots, B_L\right\} \subseteq 2^{\mathcal{C}}$ is constructed based solely on the color universe $\mathcal{C}$, strictly prior to processing the spatial coordinates of the dataset $P$. 
For each $B_i \in \mathcal{B}$, a $c$-ANN algorithm $\calA_i$ is instantiated over $P_i=\left\{p \in P \mid \chi(p) \in B_i\right\}$. 
Given a query $q$, the scheme queries all $L$ buckets to obtain a candidate set $V=\left\{\hat{p}_1, \ldots, \hat{p}_L\right\}$.
A deterministic function then selects $k$ points with distinct colors in $V$.
\end{definition} 

We next state the definition of $d$-disjunctness in the context of deterministic color bucketing, which requires each color to be ``isolated'' from all subsets of colors by at least one bucket.

\begin{definition}
A family of color subsets $\mathcal{B} = \{B_1, \dots, B_L\} \subseteq 2^{\mathcal{C}}$ (represented by its incidence matrix) is defined as $d$-disjunct (or a $(1,d)$-cover-free family) if it satisfies the following condition:
\[\forall x^* \in \mathcal{C}, \forall S^* \subset \mathcal{C} \setminus \{x^*\} \text{ such that } |S^*| = d, \exists B_i \in \mathcal{B} \text{ such that } x^* \in B_i \text{ and } B_i \cap S^* = \emptyset.\]
\end{definition}

We next show the lower bound for dataset-oblivious color bucketing schemes.
\begin{theorem}
\thmlab{thm:lb}
Assuming $k \ge 3$ and $k^2 \frac{\log N}{\log k} \le C \cdot N$ for some constant $C$, any deterministic dataset-oblivious bucketing scheme that correctly solves the $k$-diverse $c$-ANN problem for all valid datasets must evaluate $L=\Omega\left(k^2 \frac{\log N}{\log k}\right)$ buckets.
\end{theorem} 

\begin{proof}
Let $\mathcal{B}=\left\{B_1, \ldots, B_L\right\} \subseteq 2^{\mathcal{C}}$ be a fixed dataset-oblivious bucketing scheme. 
We assume for contradiction that $\mathcal{B}$ does not form a $(1, k-1)$-cover-free family (i.e., its incidence matrix is not $(k-1)$-disjunct).
Then there exists at least a color $x^* \in \mathcal{C}$ and a subset of colors $S^* \subset \mathcal{C} \backslash\left\{x^*\right\}$ of size $k-1$, such that every bucket $B_i \in \mathcal{B}$ containing $x^*$ also contains at least one color from $S^*$.

Because the scheme is dataset-oblivious, it must be universally correct for any spatial configuration of the dataset.
Then consider dataset $P$ and query $q$ satisfying the following properties:
\begin{enumerate}
\item  For every color $\gamma \in \mathcal{C}$, there is exactly one point $p_\gamma \in P$ with color $\gamma$. 
\item For all $y \in S^*, \operatorname{dist}\left(q, p_y\right)=1$.
\item $\operatorname{dist}\left(q, p_{x^*}\right)=c+\eps$, for an arbitrarily small $\eps>0$.
\item Let $K^*=S^* \cup\left\{x^*\right\}$. For all $z \in \mathcal{C} \backslash K^*, \operatorname{dist}\left(q, p_z\right)=c(c+\eps)+\eps$.
\end{enumerate}
By construction, the optimal $k$-diverse solution set for query $q$ is $K^*$, which has radius $r^*=c+\eps$. 
Since all points with colors outside of $K^*$ have distance to $q$ larger than $c(c+\eps)$, to satisfy the definition of $c$-ANNs, the algorithm is required to return exactly the $k$ points with colors in $K^*$, including $p_{x^*}$.
Therefore, there is a bucket $B_i \in \calB$ that contains $x^*$, however, by our initial non-disjunctness assumption, $B_i$ must also contain some color $y \in S^*$. 
Thus, both $p_{x^*}$ and $p_y$ belong to $P_i$. 
The minimum distance from $q$ to $P_i$ is bounded above by $\operatorname{dist}\left(q, p_y\right)=1$. 
Then any point $\widehat{p}$ returned by the $c$-ANN algorithm in bucket $B_i$ satisfies
\[\operatorname{dist}(q, \widehat{p}) \leq c \cdot \operatorname{dist}(q, P_i) \leq c \cdot 1=c.\]
Because $c<c+\eps=\operatorname{dist}\left(q, p_{x^*}\right)$, $p_{x^*}$ is not returned by bucket $B_i$. 
Thus, the algorithm fails to retrieve $x^*$ into the candidate set $V$, violating the $k$-diverse guarantee. 
Therefore, for a fixed dataset-oblivious scheme to be universally correct, the columns of the incidence matrix representing $\mathcal{B}$ must be $(k-1)$-disjunct.

By the classical extremal combinatorics theorem of \cite{d1982bounds}, the minimum universe size $L$ of a $d$-disjunct family of size $N$ is bounded below by $\Omega\left(d^2 \frac{\log N}{\log d}\right)$.
For $k \ge 3$, substituting $d=k-1$ gives us the $L=\Omega\left(k^2 \frac{\log N}{\log k}\right)$ lower bound.
Note that we implicitly assume $k^2 \frac{\log N}{\log k} \le C \cdot N$ for some constant $C$ to avoid trivial color bucketing by placing one color in each bucket.
\end{proof}

\section{Adaptive Segment Tree Bucketing}
\seclab{sec:adap}
In this section, we provide the formal discussion and analysis for our lower bound for dataset-oblivious bucketing in \secref{sec:adap:front}.
To circumvent the $\Omega(k^{2})$ barrier shown in \secref{sec:lb}, we introduce an adaptive deterministic algorithm that selectively queries buckets based on the specific spatial configuration and color assignment of the dataset, using a segment tree over the color space.
We achieve a $\log N$ overhead for space and a $k \log N$ overhead for the query time, where $N$ is the total number of colors.
This is slightly worst than the random bucketing scheme as $N \ge k$, but we achieve determinism.

\subsection{Data Structure Construction}
Assume without loss of generality that $N=2^{D}$ (otherwise, pad $\mathcal{C}$ with dummy colors that correspond to empty point sets). We construct a perfectly balanced binary tree $\mathcal{T}$ of depth $D=\lceil \log_{2}N \rceil$ as follows: 
(1) we deterministically assign one distinct color to each of the $N$ leaves;
(2) for each node $u$ in $\calT$, we include all colors associated with the leaves of the subtree rooted at $u$, and initialize a $c$-ANN data structure with the corresponding points.
We formalize this procedure in \algref{alg:det:construct}.

\begin{algorithm}[!ht]
\caption{Deterministic color bucketing construction}
\alglab{alg:det:construct}
\begin{algorithmic}[1]
\Require{Color set $\calC$ with size $N$}
\Ensure{Binary tree $\calT$ for solving the $k$-diverse nearest neighbor problem}
\State{Deterministically assign each leave of $\mathcal{T}$ a color such that the $N$ leaves of $\mathcal{T}$ are in a fixed bijection with the colors in $\mathcal{C}$}
\For{each node $u\in\mathcal{T}$}
\State{$C_{u}\gets$ the subset of colors associated with the leaves of the subtree rooted at $u$}
\State{$P_{u} \gets \{p\in P \mid c(p)\in C_{u}\}$, where $c(p)$ is the color of $p$}
\State{Initialize a standard $c$-ANN oracle $\calA_{u}$ over $P_{u}$}
\EndFor
\end{algorithmic}
\end{algorithm}

\subsection{The Query Algorithm}
At query time, the algorithm dynamically maintains a priority queue $Q$ that stores a list of pairs $(\widehat{p_u},u)$, meaning a candidate point $\widehat{p_u}$ and the node that reports it. 
We begin by adding the $c$-ANN reported by the root node.
Next, in each iteration, we first add the minimum priority element $(\widehat{p_u},u)$ in $Q$, i.e., with minimum distance to $q$, to the result set.
Then we query a group of descendent nodes of $u$ to add new candidates to $Q$ with the following requirements:
\begin{enumerate}
    \item Each node does not contain $c(p_u)$.
    \item The collection of the nodes cover all colors in $u$ except for $c(p_u)$.
\end{enumerate}
Thus, we avoid the color of $p_u$ while ensuring that all other colors in $u$ are considered.
To achieve these requirements, we look at the downward path $v_{0}=u,v_{1},...,v_{m}$ in $\mathcal{T}$ from $u$ to the leaf corresponding to $c(\widehat{p_u})$.
Note that each $v_i$ contains $c(\widehat{p_u})$, so we instead query the other child $w_{i-1}$ of $v_{i-1}$ that is not $v_{i}$ and add its reported point to $Q$, for each $i \in [m]$.
The key observation is that $\{w_i \mid i\in [m-1] \}$ forms a mutually disjoint partition of the colors in $u$ except for $c(\widehat{p_u})$.
Then under this construction, we can show that in each iteration, the remaining nodes in $Q$ forms a mutually disjoint partition of $\calC \setminus \calC_\calR$, where $\calC_\calR$ is the set of colors used by the previously extracted points from $Q$.
Therefore, in each iteration, we are able to add a new valid $c$-ANN with a distinct color to the result set, and so the algorithm terminates in $k$ iterations, using $\O{k \log N}$ queries of the $c$-ANN algorithm in total.
We formalize the above procedure in \algref{alg:query:det}.

\begin{algorithm}[!ht]
\caption{Adaptive query scheme for $k$-diverse $c$-ANN}
\alglab{alg:query:det:back}
\begin{algorithmic}[1]
\Require Query $q\in\mathcal{X}$, target diversity $k\le N$, segment tree $\mathcal{T}$
\Ensure{Solution to $k$-diverse search}
\State Initialize a priority queue $Q$, ordering pairs $(\widehat{p_u},u)$ by $dist(q,\widehat{p_u})$. Query $\calA_{\text{root}}$ to obtain a $c$-ANN $\widehat{p_{\text{root}}}$ and insert $(\widehat{p_{\text{root}}}, \text{root})$ into $Q$
\State Initialize result set $\mathcal{R}\leftarrow\emptyset$ and active colors $\calC_\calR\leftarrow\emptyset$
\While{\label{line:begin}$|\mathcal{R}|<k$}
    \State Extract the pair $(\widehat{p_u},u)$ with minimum distance to $q$ from $Q$
    \State $\mathcal{R}\leftarrow\mathcal{R}\cup\{\widehat{p_u}\}$, $\calC_\calR\leftarrow \calC_\calR\cup\{c(\widehat{p_u})\}$
    \State Let $v_{0}=u,v_{1},...,v_{m}$ be the downward path in $\mathcal{T}$ from $u$ to the leaf corresponding to $c(\widehat{p_u})$
    \For{$i=0$ to $m-1$}
        \State Let $w_{i}$ be the child of $v_{i}$ that is not $v_{i+1}$.
        \If{$P_{w_{i}}\ne\emptyset$}
        \State Query $\calA_{w_i}$ to obtain a $c$-ANN $\widehat{p_{w_i}}$ and insert $(\widehat{p_{w_i}}, w_i)$ into $Q$
        \EndIf
    \EndFor
\EndWhile
\Return $\mathcal{R}$
\end{algorithmic}
\end{algorithm}

\subsection{Analysis of \texorpdfstring{\algref{alg:query:det}}{algorithm 9}}

In this section, we analyze the correctness and complexity overheads of \algref{alg:query:det}.

\paragraph{Correctness.}
We first show the correctness of our algorithm.
To begin with, we show that the algorithm maintains an invariant after each extraction from $Q$: the nodes that have not yet been extracted from $Q$ form a mutually disjoint partition of the remaining color set.
This property is later used to show that we can always find a new valid $c$-ANN in the remaining nodes in $Q$.

\begin{lemma} 
\lemlab{lem:partition}
At the beginning of each iteration (Line~\ref{line:begin}), let $\mathcal{U} = \{w, (\widehat{p_w},w) \in Q\}$ be the set of nodes that have not yet been extracted from $Q$ and let $C_w$ denote the set of colors contained in node $w$, the collection of color sets $\{C_{w} \mid w\in\mathcal{F}\}$ forms a mutually disjoint partition of the remaining colors $\mathcal{C}\setminus \calC_\calR$, i.e., $\bigcup_{w\in\mathcal{F}} C_w = \calC \setminus \calC_\calR$ and the union is mutually disjoint. 
\end{lemma}
\begin{proof} 
We proceed by mathematical induction on the size of $\mathcal{R}$. Initially, $\mathcal{R}=\emptyset$ and $\calC_\calR=\emptyset$. 
The only queried node is the root, so $\mathcal{F}=\{\text{root}\}$.
The set $C_{\text{root}}=\mathcal{C}$ trivially partitions $\mathcal{C}\setminus\emptyset$.

Next, assume the invariant holds at the beginning of an iteration. 
We extract $(\widehat{p_u},u)$ from $Q$. 
Since $u\in Q\subseteq\mathcal{F}$, the inductive hypothesis guarantees $C_{u}\subseteq\mathcal{C}\setminus \calC_\calR$, and so we have $c(\widehat{p_u})\in C_{u}$ and $c(\widehat{p_u})\notin \calC_\calR$.
Adding $c(\widehat{p_u})$ to $\calC_\calR$ updates the excluded colors to $\calC_\calR\cup\{c(\widehat{p_u})\}$. 
This increases $|\calR|$ by $1$ and thus we proceed the induction by one step after this iteration.

Next, removing $u$ leaves a coverage gap of $C_{u}$, which is covered by the newly added points to $Q$. 
Let the path from $u$ to the leaf of $c(\widehat{p_u})$ be $v_{0}=u,v_{1},...,v_{m}$. 
Because $\mathcal{T}$ is a binary tree, for each $0\le i\le m-1$, the children of $v_{i}$ are $v_{i+1}$ and $w_{i}$. 
Thus, $C_{v_{i}}=C_{v_{i+1}}\cup C_{w_{i}}$, and $C_{v_{i+1}}\cap C_{w_{i}} = \emptyset$. 
Telescoping this identity along the path yields:
\[
C_{u}=C_{v_{m}}\cup\bigcup_{i=0}^{m-1}C_{w_{i}}=\{c(\widehat{p_u})\}\cup\bigcup_{i=0}^{m-1}C_{w_{i}}.
\]
Consequently, $\bigcup_{i=0}^{m-1}C_{w_{i}}=C_{u}\setminus\{c(\widehat{p_u})\}$, and the sets $C_{w_i}$ are mutually disjoint. 
Since we the nodes $\{w_{0},...,w_{m-1}\}$ to $\calF$ after removing $u$, the updated set $\mathcal{F}$ accurately partitions $\mathcal{C}\setminus(\calC_\calR\cup\{c(\widehat{p_u})\})$.
\end{proof}

The next lemma shows that whenever the result set contains less than $k$ points, the minimum priority element in $Q$ is a valid $c$-ANN with a new color, and thus it can be added to the result set.
\begin{lemma}
\lemlab{lem:dist:bound}
Suppose that there exists an optimal set $S^{*}\subseteq P$ of $k$ distinct colors such that $\max_{p\in S^{*}}dist(q,p)\le r^{*}$, then at the beginning of any iteration, if $|\mathcal{R}|<k$, the minimum priority element in $Q$ has distance bounded by $c\cdot r^{*}$, and moreover, its color is not in $\calC_\calR$.
\end{lemma}
\begin{proof} 
Since $|S^{*}|=k$ and $|\mathcal{R}|<k$, there exists at least one optimal color $c^*\in\chi(S^{*})\setminus \calC_\calR$. 
By \lemref{lem:partition}, the sets $\{C_{w} \mid w\in\mathcal{F}\}$ partition $\mathcal{C}\setminus \calC_\calR$. Thus, there exists exactly one node $w\in\mathcal{F}$ satisfying $c^*\in C_{w}$. 
Therefore, $\dist(q,P_w) \le r^*$ since $P_w$ contains at least one point in the optimal solution set $S^*$.
Then when the oracle $\calA_{w}$ was queried, it returns a point $\widehat{p_w}\in P_{w}$ satisfying:
\[
\dist(q,\widehat{p_w})\le c\cdot \dist(q, P_w) \le c\cdot r^{*}.\]
Therefore, $Q$ contains at least one point that has distance bounded by $c\cdot r^*$, then the minimum priority element in $Q$ is also bounded by $c\cdot r^{*}$.
In addition, by \lemref{lem:partition}, the colors covered by $\{C_{w} \mid w\in\mathcal{F}\}$ do not appear in $\calC_\calR$, so the minimum priority element has a new color.
\end{proof}

As a direct corollary of the above lemma, the next statement shows that \algref{alg:query:det} provides a valid solution.
\begin{lemma}
\lemlab{lem:det:cor}
\algref{alg:query:det} solves the $k$-diverse $c$-ANN problem in $k$ iterations.
\end{lemma}
\begin{proof} 
By \lemref{lem:dist:bound}, as long as $|\mathcal{R}|<k$, the minimum priority element in $Q$ is a valid $c$-ANN with a new color. 
Therefore we can add it to $\calR$ and proceed to the next iteration.
The algorithm will deterministically execute exactly $k$ iterations and return a $k$-diverse $c$-ANN set.
\end{proof}

\paragraph{Space usage and update time.}
We next analyze the space and time complexities of \algref{alg:query:det}.
We first upper bound the space usage

\begin{lemma} 
\lemlab{lem:det:space}
Let $S(P)$ denote the space complexity of the underlying $c$-ANN structure for a dataset $P$. 
Assuming $S(P)$ is super-additive, i.e., $S(P_1\cup P_2)\ge S(P_1)+S(P_2)$ for disjoint $P_1$ and $P_2$, \algref{alg:query:det} uses $\mathcal{O}(\log N \cdot S(P))$ space.
\end{lemma}
\begin{proof} 
For any fixed depth $l\in[0,\lceil \log_{2}N\rceil]$ in $\mathcal{T}$, the sets $C_{u}$ for nodes $u$ at depth $l$ form a disjoint partition of $\mathcal{C}$. 
Consequently, the local datasets $P_{u}$ in buckets at depth $l$ form a disjoint partition of the global dataset $P$, meaning $\bigcup_{u \text{ at depth } l} P_{u}=P$. 
Since $S$ is super-additive, the total space utilized across depth $l$ satisfies $\sum_{u}S(|P_{u}|)\le S(\sum_{u}|P_{u}|)=S(n)$. 
Summing across the $\lceil \log_{2}N\rceil+1$ depth levels, the overall space is bounded by $\mathcal{O}(\log N\cdot S(n))$.
\end{proof}

We next analyze the query time.

\begin{lemma}
\lemlab{lem:det:time}
Let $T(P)$ denote the query time of the underlying $c$-ANN structure for a dataset $P$, then \algref{alg:query:det} uses $\mathcal{O}(k\log N\cdot T(P))$ total query time.
\end{lemma}
\begin{proof}
Let $M$ be the total number of $c$-ANN queries.
By \lemref{lem:det:cor}, there is exactly $k$ iterations in \algref{alg:query:det}, and in each of the $k$ iterations, the traversal path length is bounded by the tree depth $\lceil \log_{2}N\rceil$. 
Thus, the algorithm invokes $M\le1+k\lceil \log_{2}N\rceil=\mathcal{O}(k \log N)$ queries, giving a $\O{k \log N \cdot T(P)}$ query time in total.

Next, we bound the time to maintain the priority queue $Q$.
To achieve fast query time, we can implement $Q$ as a Fibonacci heap, which supports insertions in amortized $\mathcal{O}(1)$ time and extracting the minimum element in amortized $\mathcal{O}(\log M)$ time. 
Since there are $M=\O{k\log N}$ $c$-ANN queries, the total time for insertion is $\O{k\log N}$.
In addition, since the algorithm performs at most $k$ extractions, the total extraction time is bounded by $\mathcal{O}(k \log M)=\mathcal{O}(k \log N)$. 
Therefore, the total time for maintaining $Q$ is $\O{k\log N}$, which is dominated by the query time.
\end{proof}

We note that if $T(P)$ is a concave function upper bounded by $\O{n^\rho}$ for $n = |P|$ and $\rho<1$, a finer analysis can be applied to show that the overhead is $k (\log N)^{1-\rho}$.
However, this assumption is only matched by LSH-based algorithms, which is beyond the scope of this section since we want a deterministic algorithm.
Next, we implement the graph-based algorithm in each bucket in the binary tree, achieving the following guarantees for deterministic algorithms.
\begin{theorem}[$k$-diverse $c$-ANN with general metrics]
\thmlab{thm:det:adap}
Given a set $P$ of $n$ points in a metric space $(\calX, \mathrm{dist})$ with doubling dimension $\mathbf{ddim}$ and aspect ratio $\Delta$, there is a deterministic algorithm that solves the $k$-diverse $c$-ANN problem (see \defref{def:k:div:NN}) for $c=\frac{\alpha+1}{\alpha-1}+\eps$ using 
$\O{k\log N \cdot (4\alpha)^{\mathbf{ddim}} \log_\alpha \frac{\Delta}{(\alpha-1)\eps} \log \Delta}$ query time and $\O{\log N\cdot n(4\alpha)^{\mathbf{ddim}} \log \Delta}$ space.
\end{theorem}
\begin{proof}
The correctness directly follows from \lemref{lem:det:cor}.
Then by \thmref{thm:cnn:metric}, the $c$-ANN algorithm in each bucket uses $\O{(4\alpha)^{\mathbf{ddim}} \log_\alpha \frac{\Delta}{(\alpha-1)\eps} \log \Delta}$ query time and $\O{n(4\alpha)^{\mathbf{ddim}} \log \Delta}$ space.
Note that the space complexity is super-additive, then by \lemref{lem:det:space}, the total space usage is 
\[\O{\log N \cdot S(P)}=\O{\log N\cdot n(4\alpha)^{\mathbf{ddim}} \log \Delta}.\]
In addition, by \lemref{lem:det:time}, the total query time is 
\[\O{k\log N \cdot T(P)}=\O{k \log N \cdot (4\alpha)^{\mathbf{ddim}} \log_\alpha \frac{\Delta}{(\alpha-1)\eps} \log \Delta}.\]
\end{proof}


\section{Diversity Search with Euclidean Diversity Metric}
\seclab{sec:general:diverse}
In this section, we study the $k$-diverse $c$-ANN problem under a general notion of diversity, which measures the diversity of a set $S$ by the minimum pairwise distance defined by a diversity metric $\rho$.
We state the formal definition as follows.


\begin{definition}
Let $\rho$ be a diversity metric, for a given set $S \subset \mathbb{R}^d$, its diversity is defined as the minimum pairwise distance between the points
of the set, i.e., $\mathrm{div}(S) = \min_{p,p'}\rho(p,p')$.
\end{definition}

Note that the above definition aligns with Definition A.1 in \cite{AnandIKMRSX25}, which calls a point set $S$ to be $C$-diverse if $\mathrm{div}(S) \ge C$.
Given this diversity measurement, we define the general $k$-diverse $c$-ANN problem by a bi-criteria guarantee: let $S$ be the target set of nearest neighbor with optimal diversity, then the reported set should be within $c$ times the distance between $S$ and the query, in addition, it gives a constant-factor approximation to the diversity of $S$.
We formalize this idea in the following definition.

\begin{definition}[The $k$-diverse $c$-ANN problem with diversity metric $\rho$]
\deflab{def:k:div:rho:NN}
Given a set $P$ of $n$ points in a metric space $(\calX, \mathrm{dist})$, a parameter $r > 0$, and a diverse metric $\rho$, the goal is to construct a data structure that satisfies the following: given any query $q$, if the smallest ball around $q$ which contains a set $S$ of $k$ points with $\mathrm{div}(S) = C$ has radius at most $r$, it returns a set $\widehat{S}$ of $k$ points with $\mathrm{div}(S) = \Omega(C)$ that satisfies $\max_{p \in \widehat{S}}\dist(p,q) \le cr$. 
\end{definition}

In this work, we restrict $\rho$ to be the Euclidean metric.
\cite{AbbarAIMV13} introduced an LSH-based algorithm that solves this problem using roughly $k^2 +\log n$ overhead in query time.
The algorithm maintains a list of diverse point in each LSH bucket, and then for each query, it processes in two stages: first, it selects $\O{k}$ candidates which contains the target set of diverse answers from the LSH buckets; then, it runs the GMM algorithm \cite{Gonzalez85} to select $k$ points from the candidates that gives a constant-factor approximation to the optimal diversity.
We observe that the $k^2$ bottle-neck is due to the slow processing time of the GMM algorithm, so it is our hope that if we can develop a faster algorithm to select $k$ points from the $\O{k}$ candidates, we could potentially achieve better dependency in $k$ in the update time.
Next, in \secref{sec:select:div}, we introduce a thresholding approach to select $k$ points in fast runtime, and in \secref{sec:faster:div}, we combine it with the framework in \cite{AbbarAIMV13} to achieve a query time of roughly $k^{1+\frac{1}{c^2}}$ overhead.

\subsection{Selection of Diverse Points}
\seclab{sec:select:div}

In this section, we introduce a novel algorithm to select the $k$ diverse points with fast runtime.
Our algorithm relies heavily on the $c$-ANN algorithm using LSH introduced in \secref{sec:prev:ann}.



\paragraph{Our thresholding approach.} 
Given a set $P$ of $m$ points and a subset $S \subset P$ with size $|S| = k$ and diversity $\mathrm{div}(S) = C$, our goal is to select $k$ points from $P$ with diversity $\Omega(C)$.
This can be obtained by a simple greedy algorithm with thresholding: in each round, we prune a point $p \in P$ according to some fixed order, and then add $p$ to the set of selected points $T$ if the distance between $p$ and $T$ is larger than $\frac{C}{2}$.
Note that because we assume that $P$ contains a subset $S$ of $k$ points and diversity $C$, then by the triangle inequality, there always exists some points that are visited and can be added to $T$ according to the threshold before $T$ contains $k$ points.
Thus, the greedy procedure is well-defined and output a valid solution (see detailed proof in \lemref{lem:select:div:pts}).
However, since it takes $\O{k}$ time to compute the distance between $p$ and $T$ in each round, this method requires $\O{mk} = \O{k^2}$ for $m = \O{k}$ pruning time and is prohibitively large for our goal.
To achieve efficient time, instead, we keep an LSH data structure for the selected points in $T$, then in each round, we query to data structure to obtain a $c$-ANN of $p$, which gives a constant-factor approximation to the distance between $p$ and $T$ and suffices for our purpose.
Then we only need roughly $\O{mk^{\frac{1}{c^2}}}$ total time to update the LSH data structure and query the $c$-ANN, which significantly improves on the polynomially dependency on $k$.
We present the algorithm in \algref{alg:select:div:back}.

\begin{algorithm}[!htb]
\caption{Select $k$ diverse points}
\alglab{alg:select:div:back}
\begin{algorithmic}[1]
\Require{Input dataset $P = \{p_1, \cdots, p_m\}$ with $m > k$, diversity metric $\rho$ being the Euclidean metric, constant $C$}
\Ensure{Approximate $k$ most diverse points in $P$}
\State{Let $T \gets \emptyset$}
\State{Initiate an empty LSH data structure with $L=\O{k^{\frac{1}{c^2}+o(1)} \log m}$ hash functions} \Comment{See the discussion below \defref{def:lsh}}
\For{$p \in P$}
\State{\label{line:cnn}Let $\widehat{p}$ be a $c$-ANN of $p$ in $T$} 
\If{\label{line:thres}$\rho(p,\widehat{p}) \ge \frac{C}{2}$ or $T = \emptyset$} 
\State{$T \gets T \cup \{p\}$}
\State{Hash $p$ by the $L$ hash functions in the LSH data structure}
\EndIf
\If{$|T| = k$}
\State{{\bf Return:} $T$}
\EndIf
\EndFor
\end{algorithmic}
\end{algorithm}

The next statement demonstrates that \algref{alg:select:div:back} gives a constant-factor approximation to the maximum diversity.
\begin{lemma}
\lemlab{lem:select:div:pts}
Suppose that there is a set $S \subset P$ with $\mathrm{div}(S) = C$, then \algref{alg:select:div:back} gives a set $T \subset P$ of $k$ points with $\mathrm{div}(T) \ge \frac{C}{2c}$ with high probability.
\end{lemma}
\begin{proof}
By the guarantee of $c$-ANN algorithm \cite{AndoniI08}, its failure probability is $\frac{1}{\poly(m)}$ by setting $L=\O{k^{\frac{1}{c^2}+o(1)} \log m}$.
Therefore, we condition on the success of all $c$-ANN algorithms, i.e., all $\widehat{p}$ on Line~\ref{line:cnn} of \algref{alg:select:div:back} is a $c$-ANN of $p$ in $T$, which happens with high probability by a union bound.
Fix an arbitrary $i \in [k-1]$, let $T_i$ be the set of the first $i$ points selected in \algref{alg:select:div:back}, and let $P_i$ be the set of points traversed by the algorithm before $T_i$ is obtained.
For a point $p \in P$, let $\rho(p,T_i) := \min_{p' \in T_i} \rho(p,p')$.

Now, consider an arbitrary point $u \in P$, without loss of generality, we suppose that $\rho(u,p) \le \frac{C}{2}$ for some $p \in S$.
Note that $S$ satisfies $\mathrm{div}(S) = C$, i.e., any two points in $S$ have diversity distance at least $C$, then by triangle inequality, $u$ is at least $\frac{C}{2}$ from all points in $S \backslash \{p\}$.
Thus, a $\frac{C}{2}$-radius ball centered at any arbitrary point in $P$ can cover at most one point in $S$.
Then, since $|T_i| = i$, there are at least $k-i$ points in $S \backslash T_i$ that satisfy $\rho(p,T_i) \ge \frac{C}{2}$.
Let $S'$ denote the set of such points, we assume that there is a point $s \in S'$ traversed by \algref{alg:select:div:back} before $T_i$ is obtained. 
Let $T_j$ be the set of selected points when $s$ is traversed by \algref{alg:select:div:back} where $j<i$, let $\widehat{p}$ be the selected $c$-ANN of $s$ in $T_j$ on Line~\ref{line:cnn}, then we have 
\[\rho(p,\widehat{p}) \ge \rho(p,T_j) \ge \rho(p,T_i) \ge \frac{C}{2}.\]
Therefore, $s$ satisfies the threshold on Line~\ref{line:thres} of \algref{alg:select:div:back} and is selected to $T_{j+1}$ by our algorithm construction, which contradicts $s \in S \backslash T_i$.
Then, for all $s \in S'$, we have $s \notin P_i$ and $\rho(s,T_i) \ge \frac{C}{2}$.
Thus, for all $i \in [k-1]$, there exists a $p \notin P_i$ that satisfies $\rho(p,T_i) \ge \frac{C}{2}$, so our algorithm obtains a set $T$ of $k$ points and is well-defined.

In addition, let $p_i \in P \backslash T_i$ be the point added to $T_i$ in round $i+1$ of \algref{alg:select:div:back}, then it satisfies the threshold on Line~\ref{line:thres}, i.e., $\rho(p_i,\widehat{p_i}) \ge \frac{C}{2}$, where $\widehat{p_i}$ is the selected $c$-ANN of $p_i$ in $T_i$ on Line~\ref{line:cnn}.
Then, by the guarantees of $c$-ANN, we have
\[\mathrm{div}(T_{i+1}) \ge \min\{\mathrm{div}(T_{i}), \rho(p_i,T) \} \ge \min\{\mathrm{div}(T_{i}), \frac{1}{c} \cdot \rho(p_i,\widehat{p_i}) \} \ge \min\{\mathrm{div}(T_{i}), \frac{C}{2c}\}.\]
Thus, by induction we have $\mathrm{div}(T_{k}) \ge \frac{C}{2c}$ as desired.
\end{proof}

Now, we compute the total runtime of \algref{alg:select:div:back}.
\begin{lemma}
\lemlab{lem:select:div:pts:time}
The runtime of \algref{alg:select:div:back} is $\O{dk^{\frac{1}{c^2}+o(1)}m^{1+o(1)}\log m}$.
\end{lemma}
\begin{proof}
Note that in \algref{alg:select:div:back}, there are $L=\O{k^{\frac{1}{c^2}+o(1)} \log m}$ hash functions, each using $\O{d m^{o(1)}}$ time to compute.
Then in each round, we use $\O{d m^{o(1)}L}$ time to query the $c$-ANN and hash the selected point $p$ to the $L$ buckets.
Since there are $m$ rounds, we need $\O{dk^{\frac{1}{c^2}+o(1)}m^{1+o(1)}\log m}$ time in total to maintain the data structure.
\end{proof}

We next present the formal statement of our diverse point selection algorithm.
\begin{theorem}
\thmlab{thm:k:div:selection:back}
Given a set $P$ of $m$ points in the Euclidean space $(\mathbb{R}^d, \mathrm{dist})$ such that there is a set $S \subset P$ with $\mathrm{div}(S) = C$ and aspect ratio $\Delta$, there is an algorithm that gives a set $T \subset P$ of $k$ points with $\mathrm{div}(T) \ge \frac{C}{4c}$ with high probability. The algorithm runs in $\O{dk^{\frac{1}{c^2}+o(1)}m^{1+o(1)}\log m \log \Delta}$ time.
\end{theorem}
\begin{proof}
Note that we do not know the diversity $C$ in advance, so we take geometrically increasing guesses of $C$, run \algref{alg:select:div:back} for each guess, and select the most diverse set of solutions.
Then, the correctness for each guess follows from \lemref{lem:select:div:pts} and the runtime follows from \lemref{lem:select:div:pts:time}.
Given an aspect ratio $\Delta$ of the input set, we only need to test $\log \Delta$ guesses, which gives our desired runtime.
In addition, the high probability guarantee is by union bounding across all tests.
\end{proof}

\subsection{Implication to Faster Algorithms}
\seclab{sec:faster:div}

In this section, we combine our diverse point selection algorithm with the framework in \cite{AbbarAIMV13} to achieve faster algorithms for the $k$-diverse $c$-ANN problem.
The core of their approach is the construction of $(\ell, \beta)$-robust coreset, which is formally defined as a subset $S' \subseteq S$ such that for any set of outliers $O$ with $|O| \le \ell$, the remaining set $S' \setminus O$ is a $\beta$-coreset of $S \setminus O$ for the $k$-center problem. 
During pre-processing, the algorithm stores a sequence of points in each LSH bucket that forms an $(\ell, \beta)$-robust coreset. 
At query time, the algorithm identifies a candidate set by incrementally scanning the points in each relevant LSH bucket $i$ until it finds a set $T_i$ where the number of actual outliers (points with distance $> cr$ from the query $q$) does not exceed $l$. 
By eliminating these identified outliers, the algorithm constructs an intermediate candidate set $T \setminus O$, where $T = \bigcup_i T_i$, of nearest neighbors of size $\O{k n^{\frac{1}{c}} \log k}$ that preserves diversity.
The remaining step is to run the GMM algorithm to select the $k$ points that give a constant-factor approximation of the optimal diversity (see Section 5 of \cite{AbbarAIMV13}).
Note that the bottleneck is surprisingly the GMM algorithm, replacing it with our subroutine in \secref{sec:select:div} would significantly accelerate the process.
We state the formal guarantees of the above procedure as follows:

\begin{theorem}[\cite{AbbarAIMV13}]
\thmlab{thm:k:div:rho:euclidean:set:back}
Given a set $P$ of $n$ points in the Euclidean space $(\mathbb{R}^d, \ell_2)$, there is an algorithm that selects a set $S$ with $|S|=\O{k n^{\frac{1}{c^2}} \log k}$ that contains the optimal solution to the $k$-diverse $c$-ANN problem with the Euclidean diversity metric (see \defref{def:k:div:rho:NN}). The algorithm uses $\O{k n^{\frac{1}{c^2}} \log k \cdot d}$ query time and $\O{\log k \cdot n^{1+\frac{1}{c^2}}}$ space with probability $0.99$.
\end{theorem}
\begin{proof}
This theorem follows from the proof of Lemma 8 in \cite{AbbarAIMV13}.
We note that they consider the Hamming distance in their paper and so their update time has a linear dependency in $n^{\frac{1}{c}}$ (see \thmref{thm:cnn:hamming}).
Applying the LSH construction for the Euclidean distance, we replace the $n^{\frac{1}{c}}$ factor with the $n^{\frac{1}{c^2}}$ factor in the update time (see \thmref{thm:cnn:euclidean}).
\end{proof}

Combining the subroutines in \thmref{thm:k:div:selection:back} and \thmref{thm:k:div:rho:euclidean:set:back}, we achieve the following result for $k$-diverse $c$-ANN with Euclidean diversity.

\begin{theorem}[$k$-diverse $c$-ANN in Euclidean spaces]
\thmlab{thm:k:div:rho:euclidean:back}
Given a set $P$ of $n$ points in the Euclidean space $(\mathbb{R}^d, \ell_2)$ with aspect ratio $\Delta$, there is an algorithm that solves the $k$-diverse $c$-ANN problem with diversity metric $\rho$ (see \defref{def:k:div:rho:NN}) using 
\[\O{k^{1+\frac{1}{c^2}+o(1)} (\log k)^{1+o(1)} \log n \log \Delta \cdot dn^{\frac{1}{c^2}+o(1)}}\]
query time and $\O{\log k \cdot n^{1+\frac{1}{c^2}}}$ space with probability $0.9$.
\end{theorem}
\begin{proof}
We first implement the algorithm in \thmref{thm:k:div:rho:euclidean:set:back} to select a set $S$ with $|S|=\O{k n^{\frac{1}{c^2}} \log k}$ that contains the optimal solution.
Given this set $S$, we apply our \algref{alg:select:div:back} to find a set $T \subset S$ of $k$ points with $\mathrm{div}(T) \ge \frac{C}{2c}$.
By \thmref{thm:k:div:selection:back}, the algorithm runs in $\O{dk^{\frac{1}{c^2}+o(1)}m^{1+o(1)}\log m \log \Delta}$ time, where $m$ is the size of $S$.
Since $|S|=\O{k n^{\frac{1}{c^2}} \log k}$ in our application, the total runtime is $\O{k^{1+\frac{1}{c^2}+o(1)} (\log k)^{1+o(1)} \log n \log \Delta \cdot dn^{\frac{1}{c^2}+o(1)}}$.
In addition, the space to store the data structure follows from \thmref{thm:k:div:rho:euclidean:set:back}.
\end{proof}


\section{Additional Experimental Results}
\seclab{appx:exp}

In this section, we provide additional empirical results.

%
%

\begin{figure*}[!h]
    \centering
    \begin{subfigure}{0.49\textwidth}
        \centering
        \includegraphics[width=\linewidth]{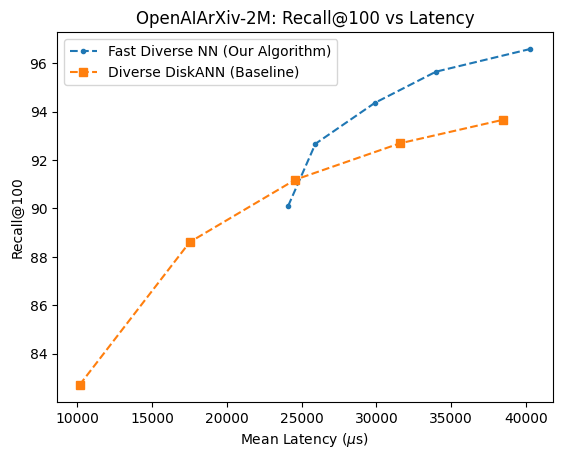}
        \subcaption{Recall-latency tradeoff}
        \figlab{fig:oai2m:lat}
    \end{subfigure}
    \hfill
    \begin{subfigure}{0.49\textwidth}
        \centering
        \includegraphics[width=\linewidth]{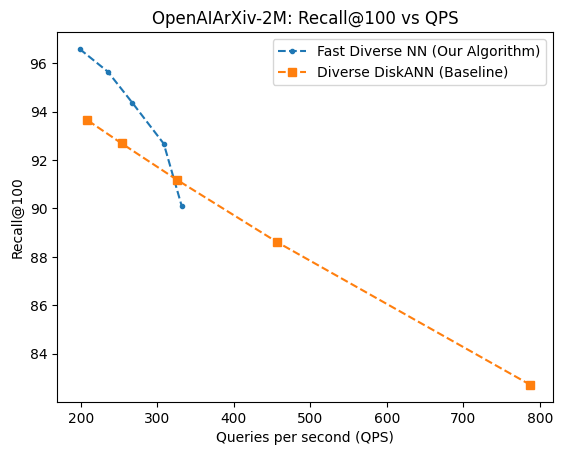}
        \subcaption{Throughput tradeoff}
        \figlab{fig:oai2m:qps}
    \end{subfigure}
    \caption{Comparisons on semi-synthetic OpenAI Embeddings Dataset}
    \figlab{fig:oai2m}
\end{figure*}

\figref{fig:oai2m:lat} and \figref{fig:oai2m:qps} compare the recall-latency and recall-QPS tradeoff of our algorithm to the baseline's on the semi-synthetic OpenAI Embeddings dataset. Our algorithm's recall is generally in a higher range (90.12 - 96.59) than the baseline's recall (82.71 - 93.66). For latency greater than 25000 $\mu$s, our algorithm produces greater recall for fixed latency. As latency increases, the gap between the algorithms widens, highlighting that our algorithm provides increased benefits for applications that require high recall. 

\begin{figure*}[!h]
    \centering
    \begin{subfigure}{0.31\textwidth}
        \centering
        \includegraphics[width=\textwidth, height=4.0cm]{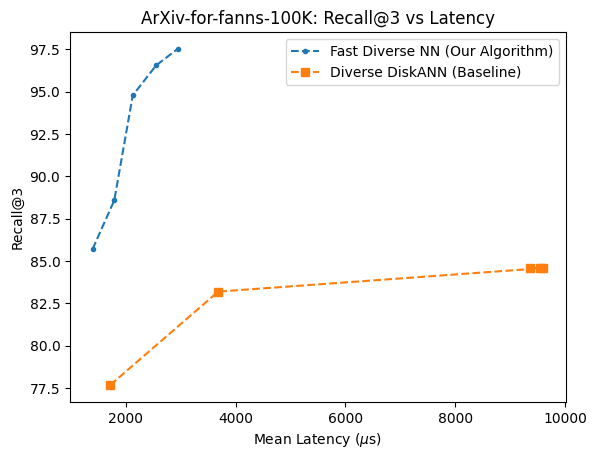}
        \subcaption{Recall-latency tradeoff}
        \figlab{fig:fanns100k:lat}
    \end{subfigure}
    \hfill
    \begin{subfigure}{0.31\textwidth}
        \centering
        \includegraphics[width=\textwidth, height=4.0cm]{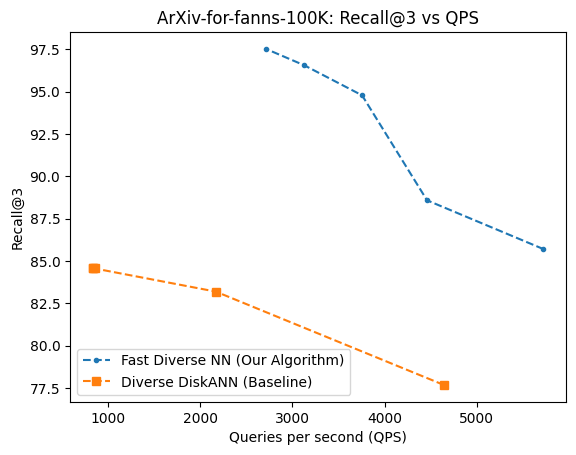}
        \subcaption{Throughput tradeoff}
        \figlab{fig:fanns100k:qps}
    \end{subfigure}
    \hfill
    \begin{subfigure}{0.31\textwidth}
        \centering
        \includegraphics[width=\textwidth, height=4.0cm]{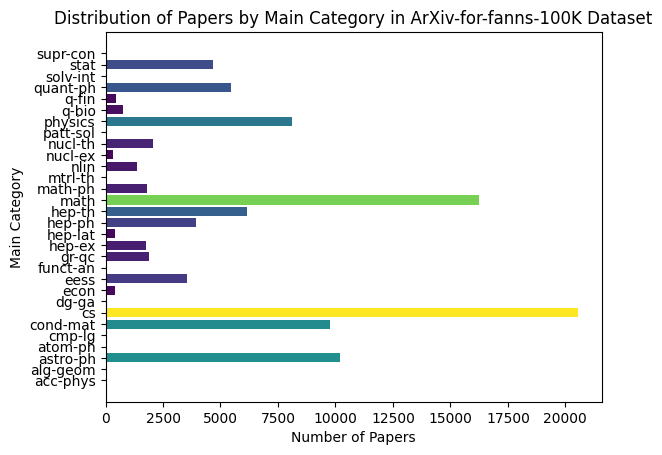}
        \caption{Paper category distribution}
        \figlab{fig:fanns100k:dist}
    \end{subfigure}
     \caption{Comparisons for real-world ArXiv-for-fanns 100K dataset}
     \figlab{fig:fanns100k}
\end{figure*}

\figref{fig:fanns100k} compares the recall-latency and recall-QPS tradeoff of our algorithm to that of the baseline on the real-word ArXiv-for-fanns dataset. Our results are consistent with the other datasets, as we have a higher recall range (85.70 - 97.52) than the baseline (77.68 - 84.58). For the same latency and QPS, our algorithm produces higher recall. As latency increases, the gap between the algorithms grows. \figref{fig:fanns100k:dist} shows the distribution of main categories or ``colors'' in the experiments involving the ArXiv-for-fanns dataset. This distribution is considerably less skewed than the synthetically generated one for the OpenAI Embeddings dataset, but our algorithm maintains its advantage.

\begin{figure*}[!h]
    \centering
    \begin{subfigure}{0.31\textwidth}
        \centering
        \includegraphics[width=\textwidth, height=4.0cm]{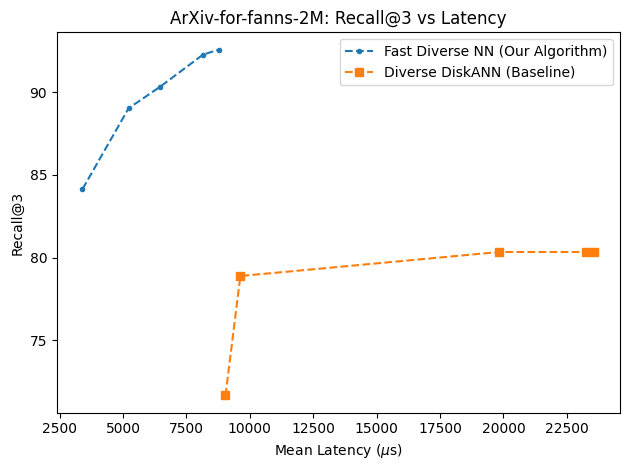}
        \subcaption{Recall-latency tradeoff}
        \figlab{fig:fanns2M:lat}
    \end{subfigure}
    \hfill
    \begin{subfigure}{0.31\textwidth}
        \centering
        \includegraphics[width=\textwidth, height=4.0cm]{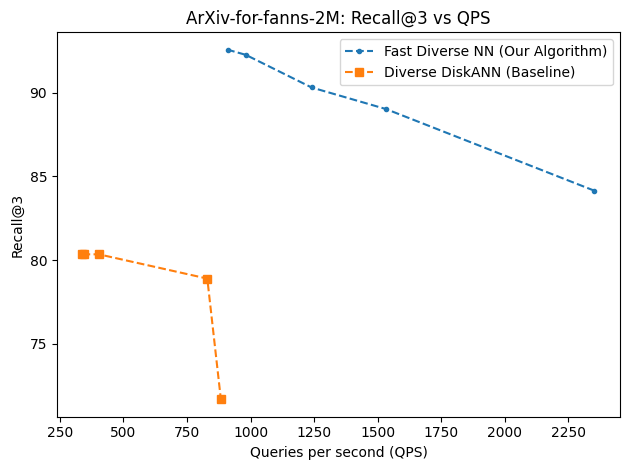}
        \subcaption{Throughput tradeoff}
        \figlab{fig:fanns2M:qps}
    \end{subfigure}
    \hfill
    \begin{subfigure}{0.31\textwidth}
        \centering
        \includegraphics[width=\textwidth, height=4.0cm]{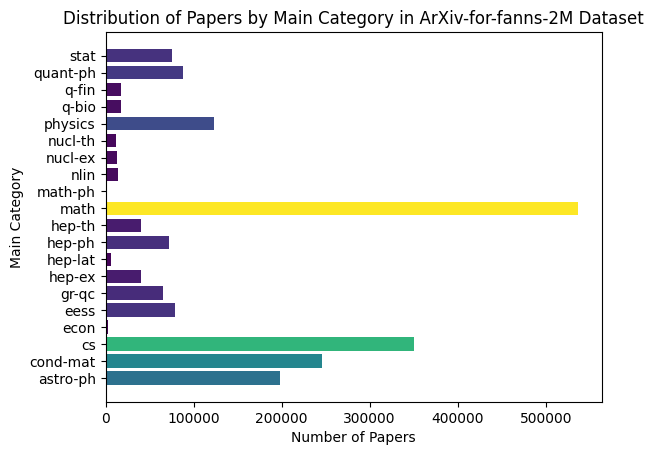}
        \caption{Paper category distribution}
        \figlab{fig:fanns2M:dist}
    \end{subfigure}
     \caption{Comparisons for real-world ArXiv-for-fanns-2M dataset}
     \figlab{fig:fanns2M}
\end{figure*}

\figref{fig:fanns2M} compares our algorithm to the baseline on the ArXiv-for-fanns dataset. While the original dataset contains approximately 2.8M total vectors, we sampled 2M vectors to feasibly obtain results. The ``colors'', i.e paper category, of the first 2M items is shown in \figref{fig:fanns2M:dist}. \figref{fig:fanns2M:lat} and \figref{fig:fanns2M:qps} compare the recall-latency and recall-QPS tradeoff of the two algorithms. Our algorithm has a higher recall range (84.14 - 92.56) than the baseline algorithm (70.71 - 80.17) and obtains higher recall for the same latency.

\tableref{tab:build} compares the index build times of the baseline to that of our algorithm. Our algorithm offer significantly improved build times for the Arxiv-for-Fanns real-world dataset and competitive performance for the LAION dataset, indicating that it may be more beneficial for real-world color distributions. Conversely, the baseline algorithm is generally faster for the semi-synthetic OpenAI Embeddings dataset.

\begin{table}
    \centering
    \begin{tabular}{l|l|l}
         Dataset & Baseline & Ours \\\hline
         Arxiv-for-Fanns 100K & 15 min. & 1 min - 3 min \\
         Arxiv-for-Fanns 2M & 5 hr. 10 min. & 48 min. - 2 hr. 8 min \\
         OpenAI Embeddings 100k & 2 min. & 4 min. - 7 min. \\
         OpenAI Embeddings 2M & 1hr 6min.& 1 hr. 35 min - 3 hr. 3 min. \\
         LAION & 36 min. & 46 min. - 49 min.
    \end{tabular}
    \vspace{2pt}
    \caption{Total Index Build Times}
    \tablelab{tab:build}
\end{table}

\begin{figure}[t]
    \centering
    \begin{tikzpicture}
        \begin{axis}[
            width=0.62\linewidth,
            height=4.5cm,
            xlabel={Number of requested neighbors $k$},
            ylabel={Mean latency (ms/query)},
            xtick={45,65,85,95},
            ytick={4,5,6,7,8},
            xmin=40, xmax=100,
            ymin=4, ymax=8,
            ymajorgrids,
            grid style={gray!25},
        ]
            \addplot[thick, mark=*] coordinates {
                (45,4.816)
                (65,6.041)
                (85,6.952)
                (95,7.510)
            };
        \end{axis}
    \end{tikzpicture}
    \caption{Mean query latency of randomized color bucketing on
    OpenAIArXiv-100 as the number of requested neighbors increases.}
    \label{fig:latency-vs-k}
\end{figure}
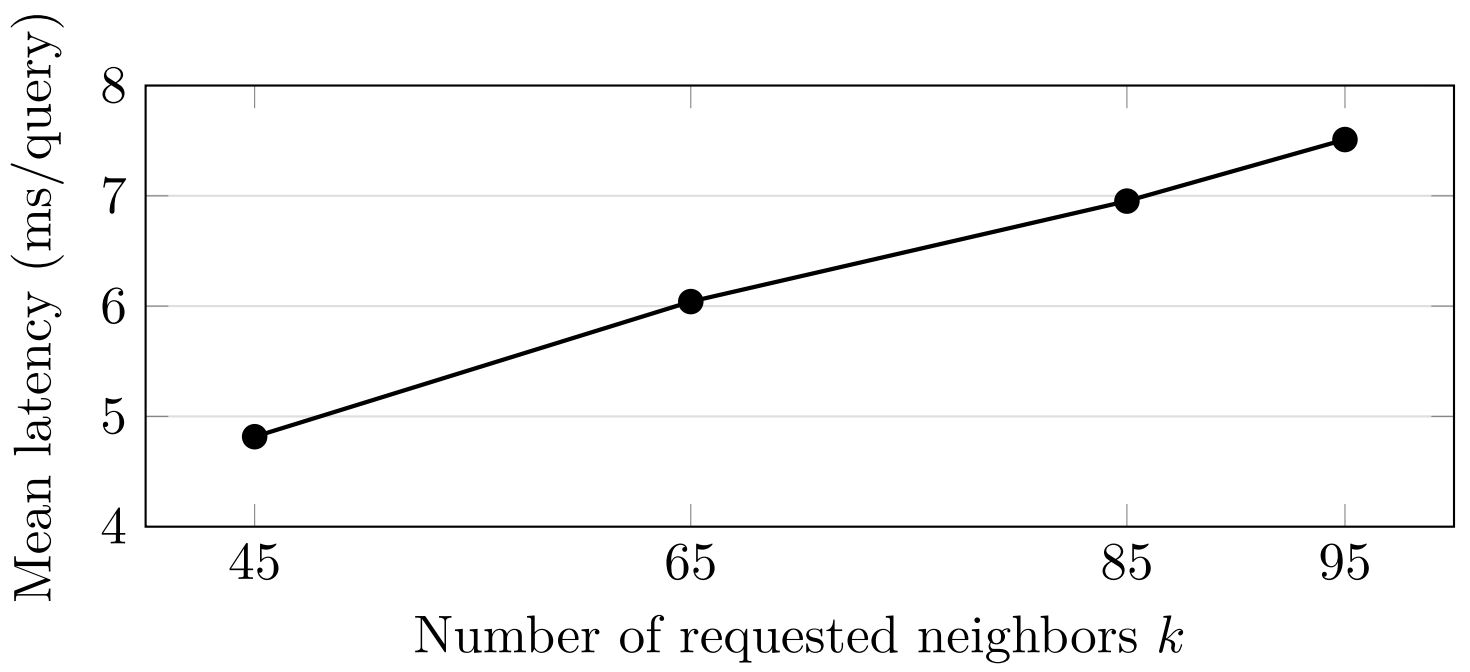

\paragraph{Query latency as a function of $k$.}
Figure~\ref{fig:latency-vs-k} shows how mean query latency varies with
the number $k$ of requested neighbors on the OpenAIArXiv-100 dataset.
Increasing $k$ from $45$ to $95$ more than doubled the number of
requested neighbors ($111.1\%$ increase), while mean latency rose
from $4.816$ to $7.510$ milliseconds per query ($55.9\%$ increase).

\paragraph{Effect of the number of buckets.}
We varied the number of index buckets $L_{\mathrm{buckets}}$ from
$600$ to $1{,}000$, keeping the maximum graph degree $R=16$ and the
search list size $L_{\mathrm{search}}=16$ fixed.
Table~\ref{tab:bucket-count-ablation} shows the resulting tradeoff.
With $800$ buckets, the algorithm attained $94.36\%$ recall at
$267.6$ queries per second. Increasing the count to $1{,}000$
gained a further $2.23$ percentage points of recall, while mean
latency rose from $29.89$ to $40.30$ milliseconds per query.

\begin{table}[t]
    \centering
    \small
    \begin{tabular}{rrrrr}
        \toprule
        Buckets & Recall (\%) & Build time (s)
                & Latency (ms/query) & Queries/s \\
        \midrule
          600 & 90.12 &  5{,}727 & 24.11 & 331.8 \\
          700 & 92.67 &  7{,}015 & 25.92 & 308.6 \\
          800 & 94.36 &  8{,}154 & 29.89 & 267.6 \\
          900 & 95.65 &  9{,}454 & 33.98 & 235.4 \\
        1{,}000 & 96.59 & 11{,}000 & 40.30 & 198.5 \\
        \bottomrule
    \end{tabular}
    \caption{Effect of the bucket count with $R=16$ and
    $L_{\mathrm{search}}=16$. Build time is the total time to
    construct all bucket indices.}
    \label{tab:bucket-count-ablation}
\end{table}


\end{document}